\documentclass[openacc]{article}%%%%where rsproca is the template name

\usepackage[british]{babel}
\usepackage{arxiv}
\usepackage{amsmath}
\usepackage{amsfonts}

\title{A Generalized Theory of Interactions - I. The General Problem}

\author{%%%% Author details
Santiago N\'u\~nez-Corrales$^{1*}$ and Eric Jakobsson$^{2\dagger}$ \\
$^{1}$Illinois Informatics and National Center for Supercomputing Applications,\\ University of Illinois at Urbana-Champaign. Urbana IL, USA.\\
$^{2}$Deparment of Physiology, Molecular and Cellular Biology and National Center for Supercomputing Applications, \\University of Illinois at Urbana-Champaign. Urbana IL, USA.\\
* Corresponding author \texttt{nunezco2@illinois.edu} \\
$\dagger$ In memoriam.
}

\begin{document}

\maketitle

%%%% Abstract text to be placed here %%%%%%%%%%%%
\begin{abstract}
Understanding realistic complex systems requires confronting significant conceptual, theoretical and experimental limitations rooted in the persistence of views that originated in the mechanics of simple moving bodies. We define the  category of complex multiscale stochastic systems as a useful device for capturing the minimally required complexity of many types of phenomena. In doing so, we provide evidence indicating that determinism, continuity and reversibility can lead to theoretical inadequacies that manifest as intractability, inaccuracies or non-representativeness when applied to complex systems. We take the view that interactions are fundamental and summarize their portrayal across many disciplines. Despite their centrality, interactions remain largely neglected as subjects of  research interest  of their own. We hypothesize that a generalized theory of interactions may help organize evidence from multiple scientific domains towards a more unified realistic view of systems.
\end{abstract}

%%%%%%%%%%%%%%%%%%%%%%%%%%%

%%%%%%%%%% Insert the texts which can accomdate on firstpage in the tag "fmtext" %%%%%

%\begin{fmtext}
\section{Introduction}

The Internet \cite{albert2000error}, bacterial communities \cite{czirok1996formation}, the global economy \cite{anderson2018economy}, ecosystems \cite{costanza1993modeling}, societies \cite{sawyer2005social}, distributed information systems \cite{barwise1997information}, international diplomatic bodies \cite{willetts2001transnational}, biomimetic nanomaterials \cite{sanchez2005biomimetism} and organizations in general \cite{dooley1997complex} exemplify systems with scientific and practical relevance characterized by dynamical complexity in their structure and function \cite{ladyman2013complex}. All of them exhibit intricate dynamics and responses coupled to their environment that impact our ability to predict their future behaviour. Elements within these systems appear simple, in stark contrast to the great variety and number of possible stimuli the system as a whole can receive \cite{ashby2011variety}. All these systems also exhibit emergent properties and self-regulation, which seem to provide resilience to small external perturbations. At a closer look, we find that their architecture depends on intricate and flexible organization patterns that are not directly dependent on external driving agents, but which lead to diverse internal microstates that work to stabilize the system as a whole against destructive influences of external driving agents and confer unity to the system \cite{ashby1969self}.

Constructing causal explanations and predictions for densely interconnected systems may or may not admit strong simplifications. Models that omit even a few elements of the problem sometimes show loss of empirical resemblance to the instance under study. Some of the most frequent simplifications involve the supposition that changes are described by linear functions or that for relevant time periods they behave \emph{almost} linearly, or that sources of noise can be safely neglected in the mathematical description. In this proposal, we provide evidence against such simplifications arguing that they are inadequate for many instances of densely interconnected systems; correspondingly, we will also describe in detail properties of these systems in an attempt to identify particular instances and formulate strategies for their modeling and simulation. Prior efforts include nonlinear science \cite{scott2003nonlinear,oono2012nonlinear} and complex systems science \cite{dent1999complexity,shalizi2006methods}. Both have made attempts to explain the structure and dynamics of these systems, but both face many challenges when used to either make empirically verified predictions or provide causal accounts of changes across temporal and spatial dimensions.

Many explorations of nonlinear science begin with viewing systems through the lens of dynamical systems theory, supposing that sets of coupled differential equations completely capture trajectories of all relevant events: a manifold contains the trajectory of variation of system states (i.e. points in their trajectories) through time, and transforming the manifold by removing time produces a phase space representation that captures the global geometry of state changes \cite{katok1997introduction}. In this view, instantaneous transformations derived from a small set of governing laws specify the changes that occur in the system. Analytic or numerical integration of these transformations --usually specified by systems of differential equations- yields values for macroscale observables that correspond to measurable properties of the real system. We discuss below several problems that exist with this approach preventing it from being effective for solving some problems in many systems.

A second view is that of networks, which has contributed in the last three decades to a vast number of discoveries and applications. A network is, briefly, a collection of entities (i.e. nodes) related in some manner (i.e. links) that determines the local and global structure of the collection, its dynamical properties or in many cases the likelihood of emergence of properties that cannot be explained by the individual action of the entities alone. Despite the success of networks in capturing essential aspects of many systems with high complexity, the picture they portray is incomplete and requires additional elements to derive causal explanation or prediction from them.

The research described here concerns itself with a third alternative aimed at  drastically improving our understanding of apparently intractable and unrelated phenomena across various systems through the lens of existing theory and practice where random noise, hierarchical modularity and irreversibility dominate the structural organization and dynamics of many interacting entities acting in some coherent and coordinated manner: we refer to them from here onwards as \emph{complex multiscale stochastic systems} (CMSS). In this introductory article we provide a definition of CMSS instances 

\section{Complex Multiscale Stochastic Systems}

In order to provide a unified account of phenomena and systems as varied as those mentioned above, we focus on understanding the properties of CMSS. We start by defining complex, multiscale stochastic systems (CMSS) as the class of systems --whether natural, artificial or informational- whose structural organization is best described by nested hierarchies of constituent systems, comprised of objects and relations subject to non-negligible and intrinsic random noise, and bearing recognizable collective identity \cite{auyang1999foundations}. Some of these systems have evolved in a biological sense, others have arisen from the combination of physical laws at different scales and, in many recent cases, these systems are being engineered by humans. Some CMSS are still in their infancy and have not reached sufficient critical connectivity for emergent phenomena to arise.

We observe that these systems exhibit four critical properties:

\begin{enumerate}
\item \textbf{Composition.} A large variety of objects and object types, driven by processes, that interact with each other in non-trivial manners.
\item \textbf{Multiscale structure.} Their structure and behavior can be described at several nested, coupled scales of aggregation by different principles and laws, yet their action remains causally connected when looked at from the outside.
\item \textbf{Stochasticity.} Random perturbations can initiate and drive sudden changes in the internal structure or dynamics of the system, leading to uncertain measurements that demand statistical treatment.
\item \textbf{Overarching laws.} CMSS obey the laws of thermodynamics: local conservation of energy (First Law) and thermodynamic irreversibility (Second Law).
\end{enumerate}

CMSS dissipate heat  \cite{seely2012fractal}, hence producing entropy, and also exhibit noise with a fractal spectrum. Living organisms \cite{schrodinger1974life} in general and the brain in particular \cite{seely2014fractal} organize in hierarchical ways and also exhibit noise with a generally fractal spectrum.. Action in CMSS follows trajectories determined by the Maximum Entropy Production (MEP) principle, which appears to create the context for the emergence of information processing \cite{dewar2003information}. Information in dissipative systems (i.e. those with friction and irreversible energy loss) often exhibit hysteresis and are characterized by one or more sources of noise for which dampening mechanisms exist at some phenomenological level. We stress here that in all prior CMSS examples our understanding of their integrated phenomenology in relation to entropy production is still at its infancy. Moving from infancy to adolescence depends on finding efficient ways to exploit general properties of underlying governing laws, and the relations between objects at different spatio-temporal scales as experimentally measured.

\subsection{Complexity}

Succinctly, a system is complex if the laws that govern the trajectory of parts do not permit a straightforward reconstruction of the trajectory of the complete system \cite{newman2011complex}, and simultaneously some type of organization is recognizable \cite{simon1977organization}. Care must be exercised not to confuse the \emph{apparent} complexity of simple dynamical systems \cite{stark2000observing} with that studied by systems captured by complex networks \cite{cilliers2001boundaries}: we wish to study systems such as clocks, made by a thousand interacting parts, instead of describing the movement of a single hinged pendulum. The complexity of the systems of interest here is beyond apparent; it manifests in their structural and dynamical properties and to some extent is captured by analogues of algorithmic complexity \cite{vitanyi1997introduction}. Moreover, the kind of complexity that captures the interest of this work is that which is simultaneously hierarchical in its structure, emergent \cite{funtowicz1994emergent} and adaptive \cite{holland1992complex} by virtue of its governing laws and composition. CMSS also manifest decentralized control strategies  \cite{siljak2011decentralized} that sustain the efficiency of internal processes by exploiting their modular structure \cite{parnas1985modular} through internal communication mechanisms. While some aspects of the energetics \cite{dyson1962statisticali,dyson1962statisticalii,dyson1962statisticaliii,dyson1963statisticaliv,mehta1963statisticalv,barabasi1999emergence} and the dynamics \cite{soodak1978homeokinetics,iberall1987physics} of systems in general have been explored through statistical physics, we lack a unified relational view of CMSS where causal relations provide information towards prediction, analysis and retrodiction based on empirical findings.

\subsection{Multiscale structure}

A complex system is also multiscale when (a) more than one valid account of same system can be given using phenomenologies that involve objects, relations and dynamics that differ widely in spatial and/or temporal scales, (b) the collective effect of governing laws and objects at one level (i.e. microscale) can be mapped to a single, much larger persistent objects at the next level (i.e. macroscale) whose properties cannot be additively explained  from objects, relations and dynamics in the former one, (c) changes in a significant\footnote{\emph{Significant} depends here on the particular system, its scales and laws proportion. What we claims is that, if a complete theory of CMSS exists, it must provide a way to compute the threshold for significance of a given current proportion.} proportion of objects in the microscale lead to distinct and mutually exclusive macrostates under certain critical conditions (i.e. phase transitions), (d) scales of structure and action can be ordered according to the amount and nature of information that can be extracted from them and (e) the architecture of the system is modular and nearly decomposable \cite{simon1991architecture}, leading to information representations \cite{majda2005information} while still determined by irreducible couplings. CMSS tend to be extremely effective negative entropy producers \cite{bar2004multiscale} that actively preserve internal stability through various mechanisms. Part of the reason seems to be connected to the numerical relations derived from their structure as described by either Zipf's law \cite{newman2005power}, power law \cite{barabasi1999emergence} or Rent's law \cite{bassett2010efficient}.

Network science has enabled the development of various methodologies capable of extracting putative structures of complex multiscale systems from data in various knowledge domains \cite{sales2007extracting}. Linking many relevant scales and including dynamics, however, remains limited in practice mainly because of the combinatorial explosion derived from keeping record of the links between different objects across scales. Partial solutions to this problem often rely on pragmatic choices that include neglecting scales, objects or relations, devising sampling strategies while preserving the general phenomenology or focusing only on one level of description to focus on a certain regime of applicability. Still, mounting research on many disciplines including physics \cite{li2001multiscale}, chemistry \cite{li2004multi}, biology \cite{engler2009multiscale,walpole2013multiscale} and socio-technical systems \cite{vespignani2009predicting} indicates that current multiscale methods face a growing number of challenges. Two specific issues are how to devise  empirically correct mappings between microscales and macroscales that capture all relevant details without sacrificing efficiency, and understanding how the underlying laws of motion at one microscale constrain or determine laws at the corresponding macroscale.

\subsection{Stochasticity}

Stochasticity refers to the presence of non-negligible noise in the dynamics of a system, often associated with nonlinearity \cite{escande1985stochasticity}. For a system to be stochastic, two types of forces must be present: drift forces that bias the motion towards some preferential direction and diffusion forces that randomize the overall trajectory of the system \cite{antonelli1977geometry}. If noise levels are small enough and fluctuations are not amplified (e.g. through multiplicative terms in the driving forces), then the system can be treated as deterministic for simplicity. However deterministic representations suffer from a deep theoretical flaw.  Because they are reversible they violate the Second Law of Thermodynamics.  Essentially they are an extension into dynamics of complex systems of the concept of the "frictionless plane" that is often introduced in introductory physics courses.  Just as one would be skeptical of the performance of a machine engineered based on the assumption of zero friction, so should one be skeptical of descriptions of complex systems based on deterministic models.  Specifically in CMSS the magnitude and frequency of fluctuations is often large enough to prompt signal propagation and autonomous reorganization through stochastic amplification \cite{volkov2003oscillatory,shibata2005noisy,li1998influence,lan2006interplay}. In such conditions, CMSS entities and dynamics appear to become organized into discrete units \cite{zeigler2000theory} --i.e. modules- that maximize internal stability in irreversible ways \cite{bennett1985role}. It is known that stability increases exponentially in complex discrete systems whose structure is described through subsystems \cite{bitsoris1977stability}, even those that contain nonlinear elements driven by time-dependent gain functions \cite{araki1981local}. The combination of these elements produces a rich landscape of possible decentralized control strategies \cite{vsiljak1978decentralized}. At present, the main challenge stochasticity introduces resides in the mathematical incompatibility of noise with the usual form of equations of motion and the  computational complexity associated to finding solutions in CMSS.

\subsection{Summary: the need for a science of CMSS structure and function}

Due to multiple scales, noisy and nonlinear dynamics, and hierarchical nature of CMSS, restricting ourselves to simplified models mostly ends up in sub-optimal solutions at best, and non-solutions often, to the problem of finding empirically relevant causal mappings across scales. In this proposal, we focus on describing various examples of how --and most important why- simplified theories and models fail to capture essential empirical aspects of CMSS. In systems with highly coupled positive and negative feedback mechanisms (e.g. \emph{wicked problems}), the density of true solutions for many problems tends to vanish compared to the size of the search space \cite{weber2017new}. However, many pressing issues in contemporary social, economic and environmental research and practice require true solutions to avoid financial, human or environmental losses.

If the principles that govern CMSS were accessible in terms of causal mappings between microscales and macroscales, gaining information about phenomena would only be a matter of computing consequences of the respective theories confident that the observables will likely be consistent with empirical measurements, including changes and their occurrence thresholds. However, finding theories that integrate nicely and universally at extreme regimes of action (e.g. quantum-classical systems) have not been found. Changes at the microscale usually involve classes of entities or relations rather than on single objects, yet many cases exist where small events (i.e. localized, unique and often very improbable) are sufficient to trigger large changes across the system \cite{holling2001understanding,albeverio2006extreme,sornette2017stock}; catastrophe theory has remained an evolving field that attempts to deal with sudden changes in systems, but its description is at the macroscale and phenomenological in nature \cite{poston2014catastrophe}.  Catastrophe theory often focuses on identifying conditions under which deterministic descriptions of dynamical systems fail, but the theory itself fails to provide an alternative description.

Our understanding of the causality of events and relations in CMSS remains limited and largely disconnected. When a system reaches a certain critical degree of complexity, it is not easy (or sometimes even possible) to accurately reconstruct the trajectory of events that lead to the current state. More importantly, even with such information, there is no guarantee that the available theories --most of them formulated as systems of differential equations or assertions about networks- will integrate nicely. The state of multiscale modeling in most areas is that of constructing a theoretical \emph{patchwork} that must be painstakingly mended to avoid breaking apart as revealed by contemporary multiscale modeling of complex materials as an example \cite{diaz2018limitations}. Changing this situation --the main objective of this proposal- is critical for the solution of many challenges of intellectual and practical importance in our century.

\section{Current problems with CMSS models across scientific theories}

The Internet, cells, ecosystems, the global economy and many other CMSS examples are characterized fundamentally by local exchanges of matter, energy or information between smaller components at all scales. Each exchange, whether evolved or designed, is constrained by governing laws of the interacting entities --or rather, the laws of their substrates which allow some notion of identity- and at the same time collections of these exchanges provide the basis for emergent constraints at critical system sizes and densities (i.e. their thermodynamic limit). We concisely review here how various interactions are addressed through various scientific fields, including how their representation and suppositions may be limited in each case. In essence, we argue that the structure of interactions is either represented explicitly without a concise explanatory picture of their dynamics or captured by detailed dynamics without any reference to its structure, becoming problematic for the purpose of having effective and efficient ways of understanding phenomena in CMSS.

Methodologically, we explore in detail the existing set of alternatives to model CMSS across various domains. First, we make explicit various assumptions about the character of dynamical equations in the context of creating effective models for CMSS. Second, we unmask a series of biases that simultaneously pervade and negatively impact the ability to reason in empirically correct ways about CMSS phenomena. Finally, we discuss the most relevant representations of interactions across the natural and social sciences.

\subsection{On the character of dynamical equations \emph{vis-\`a-vis} CMSS}

The study of fundamental relations between energy, information, time, matter and space has led to the discovery of two types of symbolic descriptions: those that pertain to concrete realizations of physical systems and those that pertain to \emph{abstract} classes of physical systems that ubiquitously constrain all concrete ones \cite{pattee2012evolving}. Correspondingly, these relations are expressed in one case as sets of dynamical equations and as physical laws in the other one. A set of dynamical equations is expected to be valid only in relation to specific settings; physical laws are more general, and the laws of thermodynamics are universal.

The usual route to arrive at specific sets of dynamical equations in contemporary physical sciences has been to depart from one of at least three starting points: its Hamiltonian \cite{lichtenberg2013regular} (i.e. a function of potential and kinetic energy), its Lagrangian \cite{peskin2018introduction} (i.e. the functional that assigns a value to a configuration of the elements of a system depending on its laws of motion) or its symmetries \cite{hojman1992new} (i.e. equivalence relations in the manifold and their consequences for motion). The existence of the three of them depends on conservation laws that ensure that no amount of motion will remain unexplained under a closed system, whether locally or globally \cite{chalmers1999making}. Conservation principles have been used  used to process trajectory data of various systems to construct nontrivial useful theories in the form of laws; generalizations which are mostly true in many instances and can be used to organize knowledge. Whether by manual or automatic means, the existence of such laws is assumed and they are sought, generally to apply to one level of system behavior.

In the context of CMSS however, characterizing the system at its outermost phenomenological level can be problematic. Choosing a level of description implies selecting certain aspects of reality, including the collection of conservation laws that apply at each level. When couplings between various scales occur, it may be either impractical to compute the consequences of laws of motion for many entities or unrealistic after simplification. More importantly, the form of the laws (and correspondingly of the dynamical equations) is of little help when stochastic fluctuations are dominant at the scale of interest. When two or more scales are included, different types of equations and laws are discontinuously patched together through approximations which lead often to semi-empirical solutions that work only for reduced regimes of action.

Yet, the existence of critical phenomena and phase transitions across a multitude of systems and scales (from the cosmological to the quantum) \cite{smolin1995cosmology} indicates that despite complex systems being nearly decomposable \cite{simon1991architecture,bassett2010efficient} they should be treated as emergent, including their governing laws. Our perception of nature as a collection of nearly decomposable systems appears to arise from our perceptual biases, the law of large numbers and the exponential decay of large deviations arising from stochastic fluctuations \cite{oono1989large,touchette2009large}. Adaptive emergence may trigger aggregation as a way to minimize action globally across the system as a means to reach regions of dynamical stability \cite{bertini2010lagrangian}. Hence, the likelihood for any arbitrary set of theories, each of them applicable only to a certain scale of action in a system --especially those that have been obtained through experimental observations at only one such scale- to integrate nicely and describe CMSS adequately (including detailed mechanisms) appears to be slim and decreases in an inversely proportional fashion with the number of scales involved.

Many, perhaps most, descriptions of dynamical systems are made through differential equations, and most often through deterministic ones. Arriving at the appropriate set of them is the bread and butter of modeling from physics to social science. In order to understand at greater depth what is involved in the choice of sets of dynamical equations, we concentrate next on their representation relative to the view of the universe as emergent from a set of fundamental laws of nature. The first two elements to be considered are concerned with how the form of the equations is derived. The third element is the determination of whether transformational pre- and post-conditions can be recovered from differential equation models. Finally we will consider physical laws as dynamical extrema under the law of large numbers.

\subsubsection{The manifold choice problem}

Let us consider the case of the classical $N$-body problem, known to be non-integrable for $N > 3$. At any time the 3D manifold contains $N$ particles, with resulting embedding of $3N$ degrees of freedom in total. The embedding itself is described as a vector product $\mathbb{R}^N \otimes \mathbb{R}^3$ of two vector spaces, that of the system of coordinates for one particle $\mathbb{R}^3$ and the collection of all particles $\mathbb{R}^N$. In general, any non-trivial motion can be described for one particle by adding its momentum to the position, leading to $\mathbb{R}^{6N}$ simultaneous variables to keep track of. Solving this computationally expensive problem is well known in computational molecular dynamics \cite{berendsen1995gromacs,phillips2005scalable}.

The input of the $N$-body problem consists of the initial positions of all the particles, the function for the pairwise (or $k$-wise) interactions between particles and the description of the energy potential surface. The more exact the model, the more computationally expensive it becomes, since the coupling of all forces (e.g. ion-induced dipole, ion-dipole, electrostatic, and van der Waals forces in molecular dynamics) in conjunction with the geometry of the objects can quickly lead to many correlations impinging on one another in small portions of the volume of interest. When the forces act at a short range (e.g. van der Waals), the problem is simplified by modeling its action with cutoff functions. In the presence of long-range forces the cutoff leads to some error which one seeks to minimize by an approximate accounting for surroundings, but some residual error is unavoidable..

Let us pause for a moment and consider various systems described by the same Hamiltonian. If we assume only classical forces are at play, all systems of interacting particles are captured by it, from very dilute gasses \cite{bhattacharya1989molecular}, liquids \cite{hess2002determining}, metallic solids \cite{fernando1989first} to core-shell nanoparticles \cite{yang2008molecular}. Four distinctively different classes of systems are captured by the same physical model, yet the complexity associated with solving their systems of equations varies drastically on each case. Given that pressing practical questions often depend on their solution, finding the smallest embedding that preserves all the features of the original system is advisable.

The problem above may be stated as that of finding an appropriate scaling relation, or a renormalization group (RG) whose effect is drastically reducing the degrees of freedom while providing a good approximation  \cite{goldenfeld2018lectures}. The scaling relation should provide three advantages: (1) reduction of the number of particles or degrees of freedom through a mean field theory, (2) identification of the critical exponents that dominate force ranges and (3) analysis schemes to reveal the emergence of structure. For the $N$-body problem with long-range correlations, a formal RG only makes sense with a few bodies (often $N \leq 4$) \cite{braaten2006universality,schmidt2010renormalization,jurgenson2011evolving}. For systems with high thermal motion, RG reduces to some form of noise across the problem domain. In both cases, information loss is guaranteed in some form of another.

On the other hand, classical mechanics provides plenty of examples of scaling relations that, though not thought of immediately as emergent, produce the same outcomes at the scale of interest and are simple to compute. For instance, take a metallic solid under translation and rotation forces. Any transformation applied to the solid is propagated across its structure to each one of its constituent atoms. Suddenly, the $\mathbb{R}^{6N}$ manifold for its dynamics reduces to $\mathbb{R}^6$, no RG required or evident at first glance. A similar case occurs for laminar flow with a suitable Reynolds coefficient, in which the dynamical equations reduce to $\mathbb{R}^2$ or even $\mathbb{R}^1$. Those are indeed extreme cases of scaling relations, the last with an effectiveness of $1/6N$; consider the implications for laminar flow of one liter of water along a pipe ($N \approx 3.346\times10^{25}$ water molecules).  It should be noted that this simple renormalization is strictly true only for a metallic solid of infinite size or laminar flow through a pipe of infinite length.

What fundamental properties of the universe as the archetype of self-organized criticality at all scales justifies such choice of degrees of freedom different from detailed descriptions of phenomenology or equational parsimony? The intuitive answer appears to be that some classes of self organization lead to collective identity \cite{nicolis1989physics}, defined as the ability of an entity to retain its structure and functions for long periods despite small perturbations. Hence, alternative models and representations of systems that may deviate from the underpinning reality become physical fictions. Such is the case of metallic crystals \cite{maradudin1963theory}: atoms are assumed to ``jiggle in place'' in a rigid lattice of (fictional) harmonic oscillators connected by springs governed by Hooke's law. A great deal of progress in crystal physics has been made in this fashion \cite{wooster2016text}, driving a large share of the present digital revolution. The springs propagate external perturbations and, at the same time, fix the global geometry of an object in place by damping propagation to adjacent atoms. The object acquires identity, and the effects of the rotation and translation of an object in $\mathbb{R}^3$ are nicely translated simultaneously (at least with respect to measurements performed at the macroscale) to the $\mathbb{R}^{6N}$ particles.

CMSS appear to lie in the middle of both purely elastic molecular collisions and metallic solids when viewed as an $N-body$ problem, yet their complexity is much higher than that in either extreme. Hierarchical modularity increases the likelihood of finding a scaling relation, but it enters immediately in tension with randomized interactions between levels of the hierarchy. In addition, and in contrast to the types of questions posed on systems modeled through the $N$-body problem, CMSS are interrogated simultaneously at multiple levels and not only at the bottom or the top of the hierarchy. Since one macroscale may be explained by a large number of equivalent microstates while events at the mesoscale remain of importance, keeping track of all entities in the hierarchy becomes a difficult task.

Concisely, except for various well-known limiting cases, finding scaling relations that yield an economic and representative manifold in hierarchically coupled systems is not a systematic process as long as Hamiltonians (or equivalent formulations) are used. Solutions are achieved by either constructing relevant abstractions  (e.g. ordered lattices) or by losing information through the reduction of degrees of freedom. At present, such essential processes for finding the most efficient and realistic manifold for CMSS instances are unknown.

\subsubsection{The functional choice problem}

Once the the manifold is chosen along with its embeddings, the next step is to determine --based on empirical observations- the most appropriate mathematical description of the phenomenon. Again, conservation laws, invariants and symmetries are the standard tools most theorists have access to for devising the equational form of the governing laws at present \cite{anco2018symmetry}. In the case of physics --and back to the $N$-body system in particular- the choice of dynamical functions occurs in a way that (a) accounts for both potential and kinetic energy present in it, (b) dictates the specific laws of motion for each body through application of differential operators to the main invariant, (c) provides coefficients that represent material aspects of the system. By fixing them, one can parameterize the equations of motion to the particular instance. Motion, in this sense, is represented as the trace in time of positions and velocities either through time series (i.e. position or velocity per every time instant) or phase diagrams (i.e. reachable states as represented by statistical properties).

As discussed above, interactions appear to be mostly hidden in dynamical descriptions of CMSS, or may be neglected. In the case of a pendulum, it is not easy to derive a picture containing interactions, yet many certainly exist. We can imagine interactions between the pendulum mechanism, the pendulum and the air (when the experiment is not carried in vacuum) and the interactions between the atoms in the rod and the bob. However, that information is unavailable, or rather is \emph{obscured} by the emphasis on position and momenta. Fourier analysis has become a standard tool \cite{bloomfield2004fourier} that uncovers the frequency of events in a system, from which interactions can be derived to some extent. Detecting the frequency of events is a starting point to the investigation of their mechanisms and internal dynamics. Both the time series and the frequency domain representation are naturally expected to be much richer and harder to unravel for CMSS.

The evident limitations of dynamical models of systems for capturing interactions requires some historical investigation about their origin. Our modern depictions of dynamical systems, and of motion in general, can be traced back to Descartes \cite{descartes19751644} and his philosophical investigations on the material conditions required for a science of motion. His analytic geometry \cite{boyer2012history} provided a foundation benefiting from both geometry and algebra as means to understand motion as a  sequence of instances, each one of them localized at a given instant of time. Representing motions as functions was only natural after this, presupposing that motion at any given future instant was predictable from motion at previous instance provided the universe contained some sort of regularity. Note that this particular definition of motion does not require any reference to interaction (or rather, a model of it) to be considered complete. In that sense, choosing the functional form to describe the motion of a dynamical system immediately refers to the selection of the appropriate number of spatial dimensions such that, at every instant, the position of an object can be located and its future position for any given time step can be known using information from the present and the past. Newton's Principia  \cite{newton1744philosophiae} adds the concept of interactions with a functional form and thus represents the pinnacle of this sort of mechanics, providing a mathematical toolkit capable of answering questions about position, velocity and momentum at any instant provided the functions were guaranteed to be continuous.

However, even at the time in which the Principia became popular, important objections were raised by Leibniz's \cite{leibniz1989specimen}. Disregarding interpersonal disputes and feuds \cite{cassirer1943newton}, two of the limitations he stated of Newton's mechanics would remain valid criticisms centuries later: (a) the mathematics of the Principia could not easily account for the material constitution of objects --or its internal motion if any- and, most importantly, (b) it disregarded completely the relation between entities and objects. The foundations of Leibniz's scientific work were placed on the putative existence of relations between objects as defined by changes in their properties, which could be identified as motion when these properties were represented in a Cartesian manner. However, the lack of impact and success of his theory of relations --in contrast to Newtonian mechanics- appears to have been largely due to his inability to articulate what interactions \emph{are} instead of what they \emph{are not} \cite{miller1988leibniz}. In that sense, Leibniz was unable to provided a computable account of motion starting from relations, and his notion of interacting bodies became confounded often with his interacting monads, which pertain to formal logic \cite{burdick1991leibniz}. As an addendum, Fourier showed that trajectories could be reversibly converted into frequencies, which when taken at infinitesimally small intervals, resembled discrete events \cite{grattan2003joseph}. The latter would be used extensively by Dirac \cite{hassani2009dirac} through his delta function to give operational meaning to interactions in quantum mechanics.

With the advances in the mathematical language of calculus and its increasing use in science and technology, the representation of motion as trajectories that are constrained by laws and given by differentiable laws of motion became the rock on which Maxwell's work on electromagnetism \cite{maxwell1904treatise} stands. However, the renaissance of atomic theory through Boltzmann's equations \cite{boltzmann1894integration,boltzmann1895certain} for the theory of gases gave way to the possibility of computing thermodynamic quantities in a straightforward fashion restricted to gas molecules that do not interact, which only holds for ideal gases. In order to include interactions, nonlinear (e.g. product) terms need to be introduced in the equations of motion, and the information that can be obtained locally is how the particular force involved (for instance, electromagnetism) would be rendered in space-time as a landscape of potential energy with valleys and peaks. The latter does not only leave the problem of describing the \emph{internal mechanics} of interactions unsolved, but also makes their numerical treatment intractable for many molecules. At present, the combinatoric explosion that arises when simulating CMSS with many scales and  classes of entities using Newtonian mechanics is a particular case of the latter problem \cite{rapaport2004art}.

Ernst Mach's \cite{mach1907science} objections to the foundations of mechanics between the XIX and XX centuries rested on the same arguments that Leibniz had raised, only in a slightly different form. Ignoring interactions, or rather the composition and properties of the relations between objects, leaves a plethora of unanswered questions and a complicated mathematical conundrum. For Mach, the ultimate form of any law of nature should be given in terms of properties of one system in relation to other systems. Einstein's special \cite{einstein1905elektrodynamik,einstein1905tragheit,einstein1906prinzip,einstein1907relativitatsprinzip} and general relativity \cite{einstein1911einfluss,einstein1912theorie,einstein1913entwurf} were non-obvious but natural extensions of Leibniz and Mach insofar as the principle of relativity was applied to frames of reference, which led to the interpretation of space-time as a dynamical entity. The objects that interact in general relativity are frames of reference and the messengers are light particles. By sending photons and experiencing internal changes, one frame of reference exerts a causal effect on another frame of reference. Although actual computations are performed by unfolding the tensor expressions into their full differential form, the most relevant representation is given by the metric tensors that indicate how energy and matter constrain space-time.  These expression are compact and appear be universal.

The development of quantum mechanics required inevitably describing interactions (i.e. measurements), and marked, with relativity, one of the two most profound revolutions in thinking about the physical world of the twentieth century \cite{jammer1989conceptual}. The necessary introduction of probability in the laws of motion, based both on wave equations and positions/momenta required several amendments to Cartesian spaces in order to cope with singularities and discontinuities \cite{gudder2014quantum} arising from intrinsic randomness and wavefunction collapse. Quantum theory showed that posterior mutual state relaxation in the form of decoherence \cite{zurek1992environment} is a fundamental property of interactions below the atomic scale. Quantum field theory \cite{itzykson2006quantum} is one of the most accurate formulations of reality in existence today, based on a model of interaction between fields that, when observed by a system, produces particles corresponding to field excitations. However, the transition functions still refer to elements of the Cartesian view and their effects are expressed as changes in the geometry of the fields rather than the changes in the entities themselves. The situation appears to be no different in modern particle physics \cite{thomson2013modern}, where Feynman diagrams are used to derive gruesome expressions posing a myriad of challenges when using the Standard Model to obtain particular solutions that explain experimental data. Although these are the two most successful theories in the physical sciences, some of their remaining limitations can still be traced thanks to Leibniz's objections.

Regardless of the theoretical background, two types of problems remain. First, the practical choice of parameters to accompany the functional descriptions is expected to be a matter of experimental verification through fitting. Nevertheless, CMSS instances pose challenges in this stage for a variety of reasons. For instance, it may be hard to differentiate parameters pertaining to two different trajectories that are spatially close at a measurement time, requiring a large number of measurements for their differentiation. It is not always possible to obtain good, repeatable experimental measurements. Second, two deeper theoretical problems arise in connection with the search for appropriate functional choices. The form of the laws must be guessed at some level with only circumstantial information from more fundamental scales, but the apparently connected structure of the universe \cite{simon1991architecture,smolin1995cosmology} and the Correspondence Principle in quantum mechanics \cite{ehrenfest1927bemerkung,bohr1976niels} suggest that physical laws are dynamical extrema at the thermodynamic limit. The form of that limit, however, can only be guessed for systems of simple particles and homogeneous interactions (if any). An open question is whether the number characterizing the thermodynamic limit for a given system (e.g. a CMSS) is system independent or is a function of the types and diversity of interactions. The second problem is the apparent independence between laws at the microscale and the organization of systems at the macroscale: if the macroscopic laws are independent from the physical substrates, then either the scales of organization cannot be truly connected, or the formulation of the laws are biased by our selection of relevant entities at the microscale \cite{gershenson2003can}. We are inclined to support the latter rather than the former, since the possible formulation of interactions as Barab\'asi networks suggests that laws appear to be topologically connected through substrates and bridge equations.

Although it is common (and correct) to think of macroscale properties as emerging from microscale objects and connections, in nature the macroscale exerts a formative influence on the microscale through adaptive evolution.  In constructed systems desired functional attributes plays the same role as adaptive evolution in constraining the nature of the microscale subsystems.

In summary, our usual choice of dynamical functions is most frequently determined by the Cartesian view of motion as trajectory, and the emphasis on their prediction and retrodiction rather by interrogations of its properties or structure. In most theories, interactions are either entirely neglected or indirectly represented by transition functions that refer to changes in fields or potentials, both contained in generalized versions of Cartesian spaces. Maintaining this view limits the possibility of understanding systems of larger complexity, either because of the combinatorial explosion of calculations or because of the inability of our methods to efficiently traverse the granular landscape of the resulting potential energy surfaces. A focus on interactions and relations between objects has brought significant understanding of the natural world on previous occasions, but interactions largely remain circumstantial, derived or implicit despite their fundamental character. Our aim in this project is to provide an alternative view of motion in CMSS where interactions put at the center or descriptions or nature and are explicitly described, investigated and operationalized.

\subsubsection{Pre- and post-conditions from dynamical equations of motion point to opaque interaction mechanisms}

The formalization of CMSS using sets of deterministic differential equations is thought to express instantaneous rates of change. Let us suppose that a given CMSS is studied using ODEs for the present argument. Take a simple ODE such as

\begin{equation}
    \label{eq.ode_example}
    \frac{d x(t)}{d t} = f(x, t)
\end{equation}

and, for the moment, remove concerns about the specific properties of $f(x, t)$. At first impression, this may be thought of as a definition only that encodes two statements: one about equality between two quantities and one about dynamics. The transformation $f$, dependent on $x$ and $t$, can be intuitively thought of as either a fixed relation in the $(t, x(t))$ manifold --more properly a graph of such instantaneous relations-- or a unique trajectory across the manifold. However, when we expand the definition of $\frac{d x(t)}{d t}$ into Eq. \ref{eq.ode_example}

\begin{equation}
    \label{eq.ode_limits}
    \lim_{\alpha \to 0} \frac{x(t + \alpha) - x(t)}{\alpha} = f(x, t)
\end{equation}

it becomes clear that, for the limit to exist, both sides of the limit must converge to $f(x,t)$ from the left and the right simultaneously. Recalling that $|\alpha| > 0$ and $|\alpha| \to 0$ in order for it to be an infinitesimal, we may think at any given moment that, for any $\alpha^- = -\alpha$ and $\alpha^+ = +\alpha$ it must be true that

\begin{equation}
    \label{eq.ode_bounded}
    \frac{x(t + \alpha^-) - x(t)}{\alpha^-} < f(x, t) < \frac{x(t + \alpha^+) - x(t)}{\alpha^+}.
\end{equation}

More generally, we may think of $\alpha$ as the limit of a Cauchy sequence $\alpha = \lim_{n \to \infty} \alpha_n$ with $\forall \alpha_i. \alpha_i \neq 0$. Hence, we may think of the left and right limits more properly as

\begin{equation}
    \label{eq.ode_cauchy}
    \frac{x(t + \alpha^-_n) - x(t)}{\alpha^-_n} < f(x, t) < \frac{x(t + \alpha^+_n) - x(t)}{\alpha^+_n}
\end{equation}

and, after reorganizing terms,

\begin{equation}
    \label{eq.ode_middlepoint}
    0 < f(x, t) - \frac{x(t + \alpha^-_n) - x(t)}{\alpha^-_n} < \frac{x(t + \alpha^+_n) - x(t)}{\alpha^+_n} - \frac{x(t + \alpha^-_n) - x(t)}{\alpha^-_n}
\end{equation}

the last term in the inequality can be interpreted as a time equivalent to the center of mass (i.e. CoD: "center of dynamics")

\begin{equation}
    \label{eq.ode_centeroftime}
    \mathrm{CoD}(\alpha, x, t) = \frac{\alpha^-_n [x(t + \alpha^+_n) - x(t)] - \alpha^+_n [x(t + \alpha^-_n) - x(t)]}{\alpha^-_n \cdot \alpha^+_n}
\end{equation}

which can be given the following interpretation: the CoD is the centroid of the interval that bounds $f(x,t)$ to the left by an event $x(t + \alpha^-_n)$ and to the right by another event $x(t + \alpha^+_n)$ for any $a_n > 0$ in the corresponding Cauchy sequence. To that extent, independent of how small $\alpha^-_n,\alpha^+_n$ become as given by approximations of the sequence, both bounding points are identified and have a definite value. It is then possible to label $\alpha^-_n$ as a pre-condition and $\alpha^+_n$ as a post-condition. Even with the instantaneous mode of action expected of Newtonian mechanics, pre-conditions and post-conditions are inescapable. In practice, this becomes more evident by the suitable choice of the integration algorithm and the integration step. One or more past points in the trajectory and estimates of one or more future points in the trajectory are used to compute the current state of the system\footnote{Pragmatically, however, equations for the current state under computation are performed for $t + \Delta t$ with $\Delta t$ being the time step.}.

What information is then contained in dynamical equations, and more importantly, what information is excluded? From the discussion above, sets of coupled deterministic differential equations do contain an implicit specification of local pre-conditions and post-conditions at every point in the manifold where the action functional is defined. We must note that, in terms of causality, trajectories are not contingent on external factors in Newtonian mechanics –-they are inevitable- but at the same time every point in the trajectory appears as if it were contingent on the (densely) immediate past. Also, these equations provide a global geometry, a field that strictly conditions the set of all possible trajectories for a given set of parameters; this does not, however, ensure that the trajectories will be sensible always or that local information suffices to predict behavior at large. Even if the equations govern the trajectory of one entity instead of the evolution of sets of entities described through proportions, there is no information whatsoever about how entities are internally structured, how their properties vary by being in contact with the medium as captured by the manifold and, in the case of coupled equations, what types of interactions lead to the coupling itself.

In summary, deterministic coupled differential equations contain a normative prescription for the action in a system by describing possible trajectories under fixed parameterizations. However, the equations themselves contain no information about mechanisms that explain the form of differential functions (validly interpreted as a transformation of state), the connection of the medium of action and the entities that explains the metric that arises under integration, and how the magnitude and frequency of couplings occurs as given by covarying quantities or proportions in a system. Intuitively, the latter framework should be considered as extremely limited to describe CMSS phenomena.

\section{Interactions: a theoretical inventory}

Interactions are necessary to understand all phenomena across the natural and social sciences. However, their representation is not transparent in terms of internal structure or processes, and in the majority of cases their presence is inferred indirectly from dynamical equations that focus on the trajectories of systems, their components or statistical correlations. Moreover, interactions in quantum mechanics and field theory are represented by unitary transformations that dictate how a prior state transforms into a posterior one without accounting for the mechanics of the change in a direct or detailed manner. A subset of social theories explicitly model the dynamics of the interactions locally, but are yet to produce an integrated understanding of the dynamics and possible macroscale configurations of social systems at large. Changes in the \emph{modes} of interaction of simpler parts of a system, however, have been experimentally shown to have aggregate or collective effects in the system as a whole. The following is a detailed description of how scientific disciplines capture interactions according to their theoretical concepts and tools.

\subsection{Statistics}

In statistics, interactions are modeled by non-additive correlations between two or more variables either explicitly through their product, or transitively by composition of other products \cite{southwood1978substantive,friedrich1982defense}. Many statistical models follow the Fisherian (i.e. frequentist) approach that asserts the statistical significance of events \cite{efron1978controversies,ziliak2008cult} by computing values of various statistics and their various moments. Its main sources of trouble are the inability to straightforwardly link statistics to observables in a system, and the need to assume normally distributed data for many of the interpretations to be correct while non-normally distributed data dominates empirical observations. Statistical models (possibly capturing the effects of interactions indirectly) often require dummy variables with no phenomenological significance for a good fit \cite{braumoeller2004hypothesis}.

Mixed models containing positive and negative regulatory effects of interactions may not lead at all to statistically significant conclusions for complex situations \cite{mcclelland1993statistical}. When the interactions are complex --beyond two-variable couplings- computational cost grows exponentially \cite{van1995multiplicative}. Various methods and heuristics are used to give structure to the statistical search space based on parsimony as a desirable property of good models \cite{austin2004bootstrap,berry2012improving}. Nevertheless, the problem of deriving mechanisms from statistical models for a given system --assuming interactions exist- in a more direct way \cite{macdonald2002incompleteness} remains open in Fisherian statistics. The needs for analysis of ever-increasing datasets in biology is suggestive of the latter \cite{wang2011statistical}. Bayesian \cite{lee2012bayesian} and non-parametric \cite{corder2014nonparametric} statistics have arisen and matured as responses to some of these limitations.

\subsection{Computing and logic}

In computer science and formal logic, interactions have been defined multiple times in relation to hardware, formal descriptions of computations, events in software systems, or their various combinations. The geometry of interactions \cite{girard1989towards} in logic attempts to capture the causal effect of sequential steps in proofs, where proofs are actions denoting observable transformations of statements grounded on an underlying operational semantics. In the classical theory of computation \cite{davis1982computability}, a Turing machine measures (i.e. interacts) with an infinite tape with cells that contain symbols: the measurement is used by a transition function to select the next state through pattern matching and its posterior selection of the associated action (e.g. read, write, move left, move right, halt). Consequently, the Turing machine is one example where interaction mechanisms in imperative programming are entirely visible thanks to the simplicity of the components and steps required to implement the robotics (i.e. automatizable manipulation) of symbolic pattern matching. In object oriented programming, the effect of a program --as formalized in sigma calculus \cite{abadi1995imperative}- is a result of the interaction between s random-access memory (RAM), a stack and a set of instructions that specify how the internal state and the computer (i.e. stack + RAM) are simultaneously modified under appropriate conditions. In constraint logic programming \cite{wallace2002constraint}, not only proofs are sequences of action that capture the interaction between denotational and operational contents of programs (e.g. the interaction between knowledge bases with rules of inference \cite{minker2014foundations}) but they are bound by the contours of internal laws (i.e. constraints) that prevent or allow only certain interactions to occur. Finally, monads were introduced in lambda calculus and functional programming languages as a means to express interactions between purely functional programs and operations that alter the machine state (i.e. the outside world) through persistent, secondary effects \cite{moggi1991notions}.

With the development of distributed computing and information systems, their interaction became a significant area of research \cite{andrews1991paradigms}. Milner's seminal Turing Award Lecture \cite{milner1993elements} recognized the need for a theory of interactions in computer systems beyond C. A. R. Hoare's foundational work con communicating sequential processes \cite{hoare1978communicating}. Simultaneously, actor-based parallel programming models for distributed systems were developed based on various empirical observations that collections of agents can achieve coordination as a result of the product of their internal (active) state and the messages being exchanged between them \cite{agha1985actors}. The expected internal complexity of agents can be derived through the law of requisite variety \cite{ashby2011variety} as well as through measures of effective complexity of systems \cite{gell1996information}: the information content of the control of a distributed system is proportional to the information required to describe its state and dynamics \cite{ranganathan2007complexity}.How interaction patterns determine the dynamics of a distributed system is a well-established research area at present \cite{tzou2007dynamics}.

Research on distributed systems has also focused on modeling and expressing interactions through programming languages \cite{agha1997abstracting,pryce1998component} and their semantics \cite{talcott1997interaction,basu2008distributed}, necessary to build and execute computer experiments that probe their effects controlled ways. Another view on distributed systems has been developed from organization theory by attempting to understand how organization patterns emerge based on which interactions are present and what the responses of the agents are \cite{fox1988organizational}, similar to what happens in human organizations. The latter has been critical to postulate and simplify the process of reasoning about collective and emergent properties of distributed systems in a succinct and intuitive manner \cite{ghenniwa2000interaction}. Establishing causality in distributed computing and information systems, despite their simplicity in comparison with other types of distributed systems in nature, has proven to be a hard problem. Lattice theory and discrete time \cite{raynal1996logical} have been used to understand collective problems in distributed information systems, which appear to strongly depend on whether interactions are synchronous or asynchronous \cite{van2008synchronous}. An interesting aspect that has been explored is how law-governed interactions shape coordination and control \cite{minsky2000law}, similar to what occurs in physical and biological systems with direct applicability to cyber physical systems \cite{rajkumar2010cyber}. In terms of theoretical advances, tangle machines \cite{carmi2014tangle,carmi2014tanglei} a model based on the intersection of interaction histories as processing.

\subsection{Social theory}

The literature describing social interactions is vast, spanning microbiology \cite{li2012quorum}, developmental biology \cite{ghabrial2006social}, neuroscience \cite{olivier1989serotonergic}, primatology \cite{byrne1989machiavellian,aureli1999heart},  human psychology \cite{lewis1981direct}, linguistics \cite{cho2000role,lakkaraju2008norm}, social science \cite{raub1990reputation,mendes2002challenge}, organization theory \cite{ishida1992organization,gasser1993organizations,bogers2017open}, economics \cite{becker1974theory,scheinkman2008social}, policy-making \cite{moffitt2001policy}, Internet-based communication \cite{huberman2008social,chen2011online}, complexity theory \cite{guttal2010social} and more recently discussions of how humans \cite{breazeal2004social,bordia2004problem,fathi2012social} and robots may coexist, to name a few. Identifying  \cite{graham1999towards,de2010identification,blume2011identification} and measuring \cite{glaeser2001measuring}  social interactions --and their effects \cite{manski2013identification}- is central to converting observations into reliable data  to either postulate new theories or to put existing ones to the test.  In contrast to the situation in the physical sciences, describing and understanding interactions (e.g. their structure, transformations, regularities) between two social agents has proven to be somewhat tractable and useful compared to efforts that have focused on obtaining governing laws. These laws are generally expected to be emergent and its form depend on the internal properties of agents and evident from various collective phenomena \cite{epstein1999agent}. Social coordination is a prime example of collective behavior that depends heavily on the properties of interactions  \cite{winograd1993categories,chwe2000communication,oullier2008social,turiel2008social}, which also appear to allow the existence of scaling laws of various types \cite{gatti2005new,bettencourt2007growth,rybski2009scaling}. To that extent, social systems constitute a prototypical CMSS example.

Given that one aim in this project is to clarify the structures of CMSS that can be interpreted simultaneously through both network and dynamical theories while remaining agnostic of particular substrates --systems in which interactions among agents appear to be crucial-, we have chosen three particular views on social systems as the most relevant descriptions of social individual and collective organization: Luhmann's systemic view of society \cite{luhmann1995social,luhmann2006system}, Berger's model of the function of communication as a form of interaction that decreases future uncertainty of encounters, \cite{berger1982language,berger1987communicating} and Pentland's conceptualization of routines as recurring patterns of interaction that simultaneously endow systems with structure and flexibility \cite{feldman2003reconceptualizing,feldman2016beyond}. Societies are complex systems in which events are coupled at multiple interconnected levels that are structured by self-organization mechanisms. These usually harness feed-back loops through coordination tokens (e.g. \textit{messengers}) that are exchanged during the interaction and encode complex transformations. These tokens, to a large extent, summarize the abstract space of state transformations as given by past and future states of the system in relation to the message being conveyed. The token therefore also encodes aspects of the internal changes in each state (e.g. relaxation, new forms of self-organization) and  causal properties of state transformations in other agents including its knowledge and actions. As ensured again by various measures of total information \cite{gell1996information} and the law of requisite variety \cite{ashby2011variety}, we can known --at least approximately- the amount of degrees of freedom exchanged between both agents by analyzing the tokens. However, the exchange is often imperfect thanks to noisy channels (i.e. environmental perturbations) or problems of interpretation (i.e. encoding mismatches), extending the effects of uncertainty to all causally connected events in the system at various degrees. Considering routines as persistent interactions introduces frequency as a relevant factor, modulated by system size and diversity.

\subsection{Biology}

Interactions are essential in biology at all scales. In ecology, the largest class of interactions are defined between the ecosystem and evolutionary rules in a feedback cycle \cite{post2009eco}. Species interact through the establishment of various relations, including predator-prey dynamics \cite{beauchamp2007predator}, parasitism \cite{combes2001parasitism} and symbiosis \cite{douglas1994symbiotic}. These relations exist even at the cellular level \cite{martin2009cell} and are complemented by chemically mediated interactions \cite{loewenstein1981junctional}. Within the environment of the cell, interactions occur in a complex biochemical environment articulated by molecular expression pathways whose activation is stochastic and threshold-dependent \cite{segal2003discovering}, and whose action is carried out by protein-protein interaction networks as determined by the affinity between various binding sites \cite{howell1992protein}. An estimate of $10^{14}$ atoms in a cell and a similar number of cells in higher organisms suffice to make numerical simulations intractable if all forces and interactions are included. However, hierarchical modularity also guarantees that approximate simulations of these systems or their components may be both tractable and remain biologically meaningful \cite{qu2011multi}. Multiscale modeling often help maintain consistency across all nested hierarchical scales without sacrificing good performance and repeatability. This network embedding can also be used to apply analogues of renormalization group theory to reduce complexity through mean-field approximations \cite{albert2002statistical} that contain scaled down quantities of entities, interactions or even hierarchical levels while preserving the main character of events and relations present in biological phenomena. Life is a prime example of a coupled, multiscale problem.  The longest relevant times scale in biology is evolutionary.  Over time scales ranging up to millions of years, selection pressures applied at the macroscale in the form of environmental change select variations at the molecular genetic level that are manifest as changes at the organismic level.

At the most abstract extreme in the research spectrum, Barab\'asi and Albert's seminal work on the emergence of scale-free networks \cite{barabasi1999emergence} has been applied widely in biology, in particular to organisms as one of the most important mesoscales in biology. These systems exhibit organized complexity \cite{kitano2002computational} arising from interactions that produce identity, robustness while remaining constrained by various types of underlying laws. Contrary to artificial complex systems, biological entities have emerged through evolution by natural selection, operating over whole ensembles of live entities. The rules of interaction follow the contours of laws that appear to emerge at more fundamental scales and are expected to define the ultimate boundaries of life itself. Cells are constantly subject a changing and complex environment whose response depends on stochastic delays driven by drift-diffusion forces, captured by Langevin equations \cite{frank2003fokker}. At the same time, phylogeny suggests that selection, repetition and variation thread the fabric of biological diversity. Understanding the massive complexity and variety observed in living organisms is one of the most important tasks in biology  \cite{woese2002evolution}. It is also one of imperfect reconstruction of interactions of extremely modular \cite{aziz2016early} and plastic \cite{guimera2005functional} systems. At present, one of the few principles that facilitates this reconstruction is Hennig's Auxiliary Principle, stating that sequence changes between genotypes are related to the molecular clocks in organisms \cite{rieppel2004language}, but more principles --and ideally, more evidence to support them- is required for a more solid scientific ground. We hypothesize that some of those new principles rest on properties and dynamics of interactions.

\subsection{Chemistry}

Interactions in chemistry at the atomic level are represented by various types of bonds, emerge from orbital structures \cite{albright2013orbital} which determine atomic valence and polarization due to electron motion. Bonding is determined by the time-dependent energy potential surface mediating interactions, described as short-range forces between atoms. Predicting interactions in chemistry at the molecular level, however, is extremely expensive. From a  generative perspective, the combinatorial problem of finding suitable reactions from sets of molecules becomes intractable \cite{balasubramanian1985applications,wieland1997combinatorics} since the space of possible molecules and interactions grows exponentially. If the set of molecules of interest is known as well as some of their interactions, predicting new ones requires accounting for the quantum character of the laws from which bonding emerges, introducing a large set of concerns.

For a given molecule, the strict computation of the state of nuclei and electrons across atomic orbitals depends simultaneously on how forces act on their position and spin \cite{woolley1977molecular}. With each new electron, computing superpositions of orbital wave functions is exponentially more expensive. Many quantum chemistry methods start from the Born-Oppenheimer approximation where (a) nuclei are removed from the computation since their mass and position are stable relative to the motion of electrons and (b) the individual electron wave functions $\psi_i(\vec{r}, \vec{s})$ are approximated as the product of a position-dependent wave function and a spin-dependent wave function independent of each other. We thus obtain the Hartree-Fock equations, which constitute the basis of most self-consistent field theories \cite{fischer1977hartree} used in quantum chemistry. Other methods such as Density Functional Theory \cite{parr1983density}) dispense with explicit representation of the electron cloud and rather operate on electron densities that vary depending on how the statistics of how electrons are spatially (and spin) correlated, and how they exchange energy changes in various situations as given by functionals dependent on the chemical species in the system. Along with many other approximations of the electronic structure of the atom, most of quantum chemistry is devoted to understanding the effects of various interactions within electron clouds on the change of the geometrical conformation of molecules, their polarization and reaction surfaces, and emitted spectra  \cite{cook2005handbook}.

Several approximations have been devised to alleviate difficulties in quantum computational chemistry methods. Stochastic Monte Carlo sampling helps decrease computational cost in all quantum chemistry methods \cite{hammond1994monte}. Moreover, semi-empirical methods are used whenever any variables can be considered as fixed and experimental data supply their value \cite{sadlej1985semi}. At a larger yet specific scale, interactions are more approximately captured by molecular reaction dynamics \cite{levine2009molecular}, which contain rules that seek to reproduce bonding behavior in simpler, more prescribed ways. Finally, molecular dynamics describes interactions at large between the position of atoms and potential energy surfaces \cite{ciccotti2014molecular} when the latter are previously known, without involving quantum phenomena. More recently, the rising field of supramolecular systems chemistry attempt to provide a better account of the (non-covalent) chemical interactions responsible for molecular recognition and self-assembly \cite{mattia2015supramolecular}.

\subsection{Classical mechanics}

Interactions in dynamical systems governed by classical mechanics are implicitly represented as non-linear terms containing two or more variables in the differential equations representation of the system trajectory. Interaction events may be often equated with either normal modes \cite{chechin1998interactions}, non-linear oscillations \cite{guckenheimer2013nonlinear}, field excitations \cite{nayfeh1995nonlinear}, or bifurcation points \cite{han1999interacting}. Interactions only arise indirectly as the underlying cause for observable deviations from additivity and linearity \cite{queller1984kin,lutz2002deviation,lobkovsky2011predictability} but the interaction mechanisms themselves causing the deviation in the trajectory remain hidden at various degrees.

When dynamical systems are simultaneously not analytically computable and sufficiently complex due to the presence of nonlinear terms and non-trivial couplings, Monte Carlo methods \cite{liu2008monte} can be used to approximate a solution by collecting independent samples with lower computational cost. Adjusting measurement density permits to approximately reconstruct nonlinearities with varying degrees of success. More refined well-known techniques remove the restriction of sample independence by introducing stochastic processes \cite{gilks1995markov,berzuini1997dynamic,snijders2002markov} in an attempt to preserve time dependencies arising during successive interactions. For spatial dependencies, a generalization exists in the form of sequential Monte Carlo sampling \cite{nemeth2015sequential}. In all Monte Carlo methods, probability models can be setup to mimic the outcome of particular interaction mechanisms at the expense of often impractically expensive computations if the systems are large and/or complex. However, the interactions themselves remain opaque.

\subsection{Quantum mechanics}

In quantum mechanics, interactions are modeled in general as interference patterns in the wave function resulting from coupled systems whose states are quantized \cite{albeverio2012solvable}. In measurement processes, one system is used to irreversibly interrogate another system, collapsing the state of one or more degrees of freedom that are superposed resulting in a particular \emph{observable} with a probability proportional to wave function amplitude. Superposition, entanglement and decoherence provide critical insights into interactions at the quantum level. Superposition entails the existence of shared global information across many possible states within a single system prior to being observed. Superposition also involves correlated histories of states prior to their measurement, a fact exploited by quantum computing architectures \cite{brukner2014quantum}. Quantum entanglement \cite{einstein1935can,bell2001einstein,horodecki2009quantum,raimond2001manipulating} corresponds to a interaction where a collective system contains a non-separable quantum state  capable of being preserved even when the system components are separated across arbitrary distances; operators applied to one part of the system impact the state as a whole immediately. Finally, quantum decoherence refers to the interaction between a quantum system and the larger environment, resulting in  fast decay of state superpositions \cite{zurek1992environment}.

In terms of representation, the evolution of a quantum system with $N$ bodies is given by a Hamiltonian operator with exponentially more information than its classical or relativistic counterparts \cite{coester1982relativistic,hunziker2000quantum}. Approximations used to compute outcomes usually contain wave functions for interacting pairs of quantum states in terms of their position, momenta or spin. In the case of coherent collective quantum phenomena such as Bose-Einstein condensation, interactions are represented as lattices of quantum harmonic oscillators whose properties depend on whether the lattice points are fermions or bosons \cite{kevrekidis2007emergent}. While the spectral nature of quantum states lends itself naturally to frequency domain analysis, the main focus remains around system trajectories.

\subsection{Quantum field theory}

In quantum field theory \cite{lancaster2014quantum}, particle interactions arise as interaction terms for excitation modes of local states of quantum fields: whenever two particles interact, a field can be inferred. Field theory has been instrumental in organizing and making sense of the diversity of particles and their interactions in high energy physics \cite{hesketh2016particle}. Feynman diagrams capture particle interactions as perturbative contributions to transition amplitudes between causally connected quantum states \cite{penco2006perturbation}. Translating these interaction diagrams into actual computations starts from obtainig rules derived from the interaction Lagrangian in quantum electrodynamics \cite{feynman1965quantum}, then finding a well-defined grounded state of the system, integrating over all possible histories for the quantum process in the field and using a Wick rotation to obtain an expression in terms of imaginary time \cite{wick1954properties}. Even when the interactions are explicit, the formal treatment of pairs of interacting field excitations remains constrained by trajectory-based representations.

\subsection{Summary: interactions in CMSS require better representations}

All prior descriptions of interactions share one of two limitations. Either the description avoids describing what happens during an interaction by concentrating on trajectories or frequencies, or specific interaction mechanisms becomes transparent at the expense of predictive or retrodictive power for system outcomes. We argue that, in order to make progress on describing complex problems with strongly-coupled scales of action towards higher realism requires constructing a theory \emph{de novo} capable of overcoming the epistemic biases and limitations derived from non-relational or partially-relational theories that do not promote interactions as first-class citizens.

\section{Conclusion}

Complex stochastic dynamical systems are ubiquitous and their understanding is necessary to find accurate and efficient solutions to frontier problems across a wide spectrum of disciplines in the natural and social sciences. Each CMSS exhibits multiple phenomenologies whose understanding remains largely incomplete and is often performed through multiple formal levels of description. These are often constructed assuming independence from one another in terms of governing laws and involved entities, are expressed through sets of deterministic, continuous, reversible dynamical equations, and are expected to integrate nicely \emph{ab initio}.

On the contrary, cosmology, biology and information theory strongly suggest simultaneously that not only the architecture of the universe is nicely integrated across all scales, but that the emergence of nearly decomposable systems appears to be one manifestation of a more general correspondence principle between a sequence of microstates associated with macrostates through bridge equations. In this view both objects and laws are emergent and dependent on the geometry defined by interactions yet in most dynamical descriptions of systems interactions are described implicitly through transformations, themselves often buried within the vanishing infinitesimals that define differential operations.

Such a state of affairs limits the ability to provide causal explanations of CMSS across multiple levels in general, resulting further in (a) the inability to predict responses of these systems at various horizons and (b) the inability to appropriately perform retrodiction by capturing all relevant contingencies. In practice, these two limitations seem to materialize as either theoretical incompatibilities between adjacent levels of description, manifold hardships to ensure appropriate formal conditions for numerical methods to yield approximately correct answers, and spurious results with respect to experimental measurements known to be rigorous. To a large extent, this is just a symptom of a deeper abyss dividing research in the physical sciences. Roughly speaking, we can identify two communities: those who insist on the preeminence of calculating \cite{kaiser2014shut}, and those that insist on the importance of working on the conceptual foundations \cite{darrigol2015shut}. Most contemporary approaches discussed here, eager to obtain results from readily observable entities fits in the first category, belongs to the first group. Our work is rather inspired by the second category.

We hypothesize that a universal solution to all the above mentioned problems exists in the form of a novel theoretical framework that focuses its attention on interactions rather than on trajectories, states or equations of motion. This generalized theory of interactions should be expressed as a differential geometry that maps manifolds of events in a CMSS as usually expressed into a 3-manifold named the \emph{interaction space} of the system that contains information about classes of interactions present in the system in a form suitable for the law of large numbers. In addition, a translation of invariants between the two classes of spaces should not only be possible, but proven to be more economic and effective in interaction space than in usual differential manifolds.

In consequence with the vast collection of differential methods, we also hypothesize that it is possible to partially recover the natural and expected variation produced by interactions currently neglected in existing differential models by (1) using spectral analysis of data from stochastic systems to obtain pure probabilistic models that capture relevant interaction scales, (2) introducing stochasticity into coupled, multiscale differential equation models (3) developing agent-based models that rigorously model interactions and (4) providing cyberinfrastructure to facilitate the gradual adoption of both the theory and associated methods.

\subsection*{Contributions}

S. Núñez-Corrales devised the theoretical formulation. E. Jakobsson edited and revised the manuscript

\subsection*{Competing interest}

Authors declare no competing interest.

\subsection*{Funding}
This work was supported by Illinois Informatics and the ACM/Intel SIGHPC Computational and Data Science Fellowship, 2017 cohort.

\subsection*{Acknowledgments}
This work is dedicated to the memory of late Prof. Eric Jakobsson, who passed away in October 2021.

%%%%%%%%%% Insert bibliography here %%%%%%%%%%%%%%

\bibliographystyle{plain}
\bibliography{references}

\begin{thebibliography}{100}

\bibitem{abadi1995imperative}
Mart{\'i}n Abadi and Luca Cardelli.
\newblock An imperative object calculus.
\newblock In {\em Colloquium on Trees in Algebra and Programming}, pages 469--485. Springer, 1995.

\bibitem{agha1985actors}
Gul~A Agha.
\newblock Actors: A model of concurrent computation in distributed systems.
\newblock Technical report, Massachusetts Institute of Technology, Cambridge Artificial Intelligence Laboratory, 1985.

\bibitem{agha1997abstracting}
Gul~A Agha.
\newblock Abstracting interaction patterns: A programming paradigm for open distributed systems.
\newblock In {\em Formal Methods for Open Object-based Distributed Systems}, pages 135--153. Springer, 1997.

\bibitem{albert2002statistical}
R{\'e}ka Albert and Albert-L{\'a}szl{\'o} Barab{\'a}si.
\newblock Statistical mechanics of complex networks.
\newblock {\em Reviews of modern physics}, 74(1):47, 2002.

\bibitem{albert2000error}
R{\'e}ka Albert, Hawoong Jeong, and Albert-L{\'a}szl{\'o} Barab{\'a}si.
\newblock Error and attack tolerance of complex networks.
\newblock {\em nature}, 406(6794):378, 2000.

\bibitem{albeverio2012solvable}
Sergio Albeverio, Friedrich Gesztesy, Raphael Hoegh-Krohn, and Helge Holden.
\newblock {\em Solvable models in quantum mechanics}.
\newblock Springer Science \& Business Media, 2012.

\bibitem{albeverio2006extreme}
Sergio Albeverio, Volker Jentsch, and Holger Kantz.
\newblock {\em Extreme events in nature and society}.
\newblock Springer Science \& Business Media, 2006.

\bibitem{albright2013orbital}
Thomas~A Albright, Jeremy~K Burdett, and Myung-Hwan Whangbo.
\newblock {\em Orbital interactions in chemistry}.
\newblock John Wiley \& Sons, 2013.

\bibitem{anco2018symmetry}
Stephen~C Anco and Abdul~H Kara.
\newblock Symmetry-invariant conservation laws of partial differential equations.
\newblock {\em European Journal of Applied Mathematics}, 29(1):78--117, 2018.

\bibitem{anderson2018economy}
Philip~W Anderson.
\newblock {\em The economy as an evolving complex system}.
\newblock CRC Press, 2018.

\bibitem{andrews1991paradigms}
Gregory~R Andrews.
\newblock Paradigms for process interaction in distributed programs.
\newblock {\em ACM Computing Surveys (CSUR)}, 23(1):49--90, 1991.

\bibitem{antonelli1977geometry}
Peter~L Antonelli and Curtis Strobeck.
\newblock The geometry of random drift i. stochastic distance and diffusion.
\newblock {\em Advances in Applied Probability}, 9(2):238--249, 1977.

\bibitem{araki1981local}
Mituhiko Araki and Toshio Kato.
\newblock Local stability and stability region of discrete-time composite systems.
\newblock {\em Transactions of the Society of Instrument and Control Engineers}, 17(1):23--28, 1981.

\bibitem{ashby1969self}
W~Ross Ashby.
\newblock Self-regulation and requisite variety.
\newblock {\em Systems thinking}, pages 105--124, 1969.

\bibitem{ashby2011variety}
W~Ross Ashby and Jeffrey Goldstein.
\newblock Variety, constraint, and the law of requisite variety.
\newblock {\em Emergence: Complexity and Organization}, 13(1/2):190, 2011.

\bibitem{aureli1999heart}
Filippo Aureli, Stephanie~D Preston, and Frans de~Waal.
\newblock Heart rate responses to social interactions in free-moving rhesus macaques (macaca mulatta): a pilot study.
\newblock {\em Journal of comparative psychology}, 113(1):59, 1999.

\bibitem{austin2004bootstrap}
Peter~C Austin and Jack~V Tu.
\newblock Bootstrap methods for developing predictive models.
\newblock {\em The American Statistician}, 58(2):131--137, 2004.

\bibitem{auyang1999foundations}
Sunny~Y Auyang.
\newblock {\em Foundations of complex-system theories: in economics, evolutionary biology, and statistical physics}.
\newblock Cambridge University Press, 1999.

\bibitem{aziz2016early}
M~Fayez Aziz, Kelsey Caetano-Anoll{\'e}s, and Gustavo Caetano-Anoll{\'e}s.
\newblock The early history and emergence of molecular functions and modular scale-free network behavior.
\newblock {\em Scientific reports}, 6:25058, 2016.

\bibitem{balasubramanian1985applications}
K~Balasubramanian.
\newblock Applications of combinatorics and graph theory to spectroscopy and quantum chemistry.
\newblock {\em Chemical Reviews}, 85(6):599--618, 1985.

\bibitem{bar2004multiscale}
Yaneer Bar-Yam.
\newblock Multiscale complexity/entropy.
\newblock {\em Advances in Complex Systems}, 7(01):47--63, 2004.

\bibitem{barabasi1999emergence}
Albert-L{\'a}szl{\'o} Barab{\'a}si and R{\'e}ka Albert.
\newblock Emergence of scaling in random networks.
\newblock {\em science}, 286(5439):509--512, 1999.

\bibitem{barwise1997information}
Jon Barwise and Jerry Seligman.
\newblock {\em Information flow: the logic of distributed systems}, volume~44.
\newblock Cambridge University Press, 1997.

\bibitem{bassett2010efficient}
Danielle~S Bassett, Daniel~L Greenfield, Andreas Meyer-Lindenberg, Daniel~R Weinberger, Simon~W Moore, and Edward~T Bullmore.
\newblock Efficient physical embedding of topologically complex information processing networks in brains and computer circuits.
\newblock {\em PLoS computational biology}, 6(4), 2010.

\bibitem{basu2008distributed}
Ananda Basu, Philippe Bidinger, Marius Bozga, and Joseph Sifakis.
\newblock Distributed semantics and implementation for systems with interaction and priority.
\newblock In {\em International Conference on Formal Techniques for Networked and Distributed Systems}, pages 116--133. Springer, 2008.

\bibitem{beauchamp2007predator}
David~A Beauchamp, D~Wahl, and Brett~M Johnson.
\newblock Predator-prey interactions.
\newblock {\em Analysis and interpretation of freshwater fisheries data. American Fisheries Society, Bethesda, Maryland}, pages 765--842, 2007.

\bibitem{becker1974theory}
Gary~S Becker.
\newblock A theory of social interactions.
\newblock {\em Journal of political economy}, 82(6):1063--1093, 1974.

\bibitem{bell2001einstein}
John~S Bell.
\newblock Einstein-podolsky-rosen experiments.
\newblock In {\em John S Bell on the Foundations of Quantum Mechanics}, pages 74--83. World Scientific, 2001.

\bibitem{bennett1985role}
Charles~H Bennett and Geoffrey Grinstein.
\newblock Role of irreversibility in stabilizing complex and nonergodic behavior in locally interacting discrete systems.
\newblock {\em Physical review letters}, 55(7):657, 1985.

\bibitem{berendsen1995gromacs}
Herman~JC Berendsen, David van~der Spoel, and Rudi van Drunen.
\newblock Gromacs: a message-passing parallel molecular dynamics implementation.
\newblock {\em Computer Physics Communications}, 91(1-3):43--56, 1995.

\bibitem{berger1987communicating}
Charles~R Berger.
\newblock Communicating under uncertainty.
\newblock 1987.

\bibitem{berger1982language}
Charles~R Berger and James~J Bradac.
\newblock {\em Language and social knowledge: Uncertainty in interpersonal relations}, volume~2.
\newblock Hodder Education, 1982.

\bibitem{berry2012improving}
William~D Berry, Matt Golder, and Daniel Milton.
\newblock Improving tests of theories positing interaction.
\newblock {\em The Journal of Politics}, 74(3):653--671, 2012.

\bibitem{bertini2010lagrangian}
Lorenzo Bertini, Alberto De~Sole, Davide Gabrielli, Giovanni Jona-Lasinio, and Claudio Landim.
\newblock Lagrangian phase transitions in nonequilibrium thermodynamic systems.
\newblock {\em Journal of Statistical Mechanics: Theory and Experiment}, 2010(11):L11001, 2010.

\bibitem{berzuini1997dynamic}
Carlo Berzuini, Nicola~G Best, Walter~R Gilks, and Cristiana Larizza.
\newblock Dynamic conditional independence models and markov chain monte carlo methods.
\newblock {\em Journal of the American Statistical Association}, 92(440):1403--1412, 1997.

\bibitem{bettencourt2007growth}
Lu{\'i}s~MA Bettencourt, Jos{\'e} Lobo, Dirk Helbing, Christian K{\"u}hnert, and Geoffrey~B West.
\newblock Growth, innovation, scaling, and the pace of life in cities.
\newblock {\em Proceedings of the national academy of sciences}, 104(17):7301--7306, 2007.

\bibitem{bhattacharya1989molecular}
DK~Bhattacharya and GC~Lie.
\newblock Molecular-dynamics simulations of nonequilibrium heat and momentum transport in very dilute gases.
\newblock {\em Physical review letters}, 62(8):897, 1989.

\bibitem{bitsoris1977stability}
G~Bitsoris and Ch~Burgat.
\newblock Stability analysis of complex discrete systems with locally and globally stable subsystems.
\newblock {\em International Journal of Control}, 25(3):413--424, 1977.

\bibitem{bloomfield2004fourier}
Peter Bloomfield.
\newblock {\em Fourier analysis of time series: an introduction}.
\newblock John Wiley \& Sons, 2004.

\bibitem{blume2011identification}
Lawrence~E Blume, William~A Brock, Steven~N Durlauf, and Yannis~M Ioannides.
\newblock Identification of social interactions.
\newblock In {\em Handbook of social economics}, volume~1, pages 853--964. Elsevier, 2011.

\bibitem{bogers2017open}
Marcel Bogers, Ann-Kristin Zobel, Allan Afuah, Esteve Almirall, Sabine Brunswicker, Linus Dahlander, Lars Frederiksen, Annabelle Gawer, Marc Gruber, Stefan Haefliger, et~al.
\newblock The open innovation research landscape: Established perspectives and emerging themes across different levels of analysis.
\newblock {\em Industry and Innovation}, 24(1):8--40, 2017.

\bibitem{boltzmann1894integration}
Ludwig Boltzmann.
\newblock Zur integration der diffusionsgleichung bei variabeln diffusionscoefficienten.
\newblock {\em Annalen der Physik}, 289(13):959--964, 1894.

\bibitem{boltzmann1895certain}
Ludwig Boltzmann.
\newblock On certain questions of the theory of gases.
\newblock {\em Nature}, 51(1322):413, 1895.

\bibitem{bordia2004problem}
Prashant Bordia and Nicholas DiFonzo.
\newblock Problem solving in social interactions on the internet: Rumor as social cognition.
\newblock {\em Social Psychology Quarterly}, 67(1):33--49, 2004.

\bibitem{boyer2012history}
Carl~B Boyer.
\newblock {\em History of analytic geometry}.
\newblock Courier Corporation, 2012.

\bibitem{braaten2006universality}
Eric Braaten and H-W Hammer.
\newblock Universality in few-body systems with large scattering length.
\newblock {\em Physics Reports}, 428(5-6):259--390, 2006.

\bibitem{braumoeller2004hypothesis}
Bear~F Braumoeller.
\newblock Hypothesis testing and multiplicative interaction terms.
\newblock {\em International organization}, 58(4):807--820, 2004.

\bibitem{breazeal2004social}
Cynthia Breazeal.
\newblock Social interactions in hri: the robot view.
\newblock {\em IEEE Transactions on Systems, Man, and Cybernetics, Part C (Applications and Reviews)}, 34(2):181--186, 2004.

\bibitem{brukner2014quantum}
{\v{C}}aslav Brukner.
\newblock Quantum causality.
\newblock {\em Nature Physics}, 10(4):259, 2014.

\bibitem{burdick1991leibniz}
Howard Burdick.
\newblock What was leibniz's problem about relations?
\newblock {\em Synthese}, 88(1):1--13, 1991.

\bibitem{byrne1989machiavellian}
Richard Byrne and Andrew Whiten.
\newblock Machiavellian intelligence: social expertise and the evolution of intellect in monkeys, apes, and humans (oxford science publications).
\newblock 1989.

\bibitem{carmi2014tangle}
Avishy~Y Carmi and Daniel Moskovich.
\newblock Tangle machines.
\newblock {\em arXiv preprint arXiv:1404.2862}, 2014.

\bibitem{carmi2014tanglei}
Avishy~Y Carmi and Daniel Moskovich.
\newblock Tangle machines ii: Invariants.
\newblock {\em arXiv preprint arXiv:1404.2863}, 2014.

\bibitem{cassirer1943newton}
Ernst Cassirer.
\newblock Newton and leibniz.
\newblock {\em The Philosophical Review}, 52(4):366--391, 1943.

\bibitem{chalmers1999making}
Alan Chalmers.
\newblock Making sense of laws of physics.
\newblock In {\em Causation and laws of nature}, pages 3--16. Springer, 1999.

\bibitem{chechin1998interactions}
GM~Chechin and VP~Sakhnenko.
\newblock Interactions between normal modes in nonlinear dynamical systems with discrete symmetry. exact results.
\newblock {\em Physica D: Nonlinear Phenomena}, 117(1-4):43--76, 1998.

\bibitem{chen2011online}
Yubo Chen, Qi~Wang, and Jinhong Xie.
\newblock Online social interactions: A natural experiment on word of mouth versus observational learning.
\newblock {\em Journal of marketing research}, 48(2):238--254, 2011.

\bibitem{cho2000role}
Grace Cho.
\newblock The role of heritage language in social interactions and relationships: Reflections from a language minority group.
\newblock {\em Bilingual Research Journal}, 24(4):369--384, 2000.

\bibitem{chwe2000communication}
Michael Suk-Young Chwe.
\newblock Communication and coordination in social networks.
\newblock {\em The Review of Economic Studies}, 67(1):1--16, 2000.

\bibitem{ciccotti2014molecular}
Giovanni Ciccotti, Mauro Ferrario, Christof Schuette, et~al.
\newblock Molecular dynamics simulation.
\newblock {\em Entropy}, 16:233, 2014.

\bibitem{cilliers2001boundaries}
Paul Cilliers.
\newblock Boundaries, hierarchies and networks in complex systems.
\newblock {\em International Journal of Innovation Management}, 5(02):135--147, 2001.

\bibitem{coester1982relativistic}
Fritz Coester and Wayne~Nicholas Polyzou.
\newblock Relativistic quantum mechanics of particles with direct interactions.
\newblock {\em Physical Review D}, 26(6):1348, 1982.

\bibitem{combes2001parasitism}
Claude Combes.
\newblock {\em Parasitism: the ecology and evolution of intimate interactions}.
\newblock University of Chicago Press, 2001.

\bibitem{cook2005handbook}
David~B Cook.
\newblock {\em Handbook of computational quantum chemistry}.
\newblock Courier Corporation, 2005.

\bibitem{corder2014nonparametric}
Gregory~W Corder and Dale~I Foreman.
\newblock {\em Nonparametric statistics: A step-by-step approach}.
\newblock John Wiley \& Sons, 2014.

\bibitem{costanza1993modeling}
Robert Costanza, Lisa Wainger, Carl Folke, and Karl-G{\"o}ran M{\"a}ler.
\newblock Modeling complex ecological economic systems: toward an evolutionary, dynamic understanding of people and nature.
\newblock In {\em Ecosystem Management}, pages 148--163. Springer, 1993.

\bibitem{czirok1996formation}
Andr{\'a}s Czir{\'o}k, Eshel Ben-Jacob, Inon Cohen, and Tam{\'a}s Vicsek.
\newblock Formation of complex bacterial colonies via self-generated vortices.
\newblock {\em Physical Review E}, 54(2):1791, 1996.

\bibitem{darrigol2015shut}
Olivier Darrigol.
\newblock 'shut up and contemplate!': Lucien hardy's reasonable axioms for quantum theory.
\newblock {\em Studies in History and Philosophy of Science Part B: Studies in History and Philosophy of Modern Physics}, 52:328--342, 2015.

\bibitem{davis1982computability}
Martin Davis.
\newblock {\em Computability \& unsolvability}.
\newblock Courier Corporation, 1982.

\bibitem{de2010identification}
Giacomo De~Giorgi, Michele Pellizzari, and Silvia Redaelli.
\newblock Identification of social interactions through partially overlapping peer groups.
\newblock {\em American Economic Journal: Applied Economics}, 2(2):241--75, 2010.

\bibitem{dent1999complexity}
Eric~B Dent.
\newblock Complexity science: A worldview shift.
\newblock {\em Emergence}, 1(4):5--19, 1999.

\bibitem{descartes19751644}
Ren{\'e} Descartes.
\newblock 1644 principia philosophiae.
\newblock {\em J. Cottingham, R. Stoothoff and D. Murdoch (trans.),``Principles of Philosophy'', The Philosophical Writings of Descartes}, 1:177--291, 1975.

\bibitem{dewar2003information}
Roderick Dewar.
\newblock Information theory explanation of the fluctuation theorem, maximum entropy production and self-organized criticality in non-equilibrium stationary states.
\newblock {\em Journal of Physics A: Mathematical and General}, 36(3):631, 2003.

\bibitem{diaz2018limitations}
Adrian Diaz, David McDowell, and Youping Chen.
\newblock The limitations and successes of concurrent dynamic multiscale modeling methods at the mesoscale.
\newblock In {\em Generalized Models and Non-classical Approaches in Complex Materials 2}, pages 55--77. Springer, 2018.

\bibitem{dooley1997complex}
Kevin~J Dooley.
\newblock A complex adaptive systems model of organization change.
\newblock {\em Nonlinear dynamics, psychology, and life sciences}, 1(1):69--97, 1997.

\bibitem{douglas1994symbiotic}
AE~Douglas.
\newblock {\em Symbiotic Interactions: Oxford Science Pub-lications}.
\newblock Oxford University Press, Oxford, 1994.

\bibitem{dyson1962statisticali}
Freeman~J Dyson.
\newblock Statistical theory of the energy levels of complex systems. i.
\newblock {\em Journal of Mathematical Physics}, 3(1):140--156, 1962.

\bibitem{dyson1962statisticalii}
Freeman~J Dyson.
\newblock Statistical theory of the energy levels of complex systems. ii.
\newblock {\em Journal of Mathematical Physics}, 3(1):157--165, 1962.

\bibitem{dyson1962statisticaliii}
Freeman~J Dyson.
\newblock Statistical theory of the energy levels of complex systems. iii.
\newblock {\em Journal of Mathematical Physics}, 3(1):166--175, 1962.

\bibitem{dyson1963statisticaliv}
Freeman~J Dyson and Madan~Lal Mehta.
\newblock Statistical theory of the energy levels of complex systems. iv.
\newblock {\em Journal of Mathematical Physics}, 4(5):701--712, 1963.

\bibitem{efron1978controversies}
Bradley Efron.
\newblock Controversies in the foundations of statistics.
\newblock {\em The American Mathematical Monthly}, 85(4):231--246, 1978.

\bibitem{ehrenfest1927bemerkung}
Paul Ehrenfest.
\newblock Bemerkung {\"u}ber die angen{\"a}herte g{\"u}ltigkeit der klassischen mechanik innerhalb der quantenmechanik.
\newblock {\em Zeitschrift f{\"u}r Physik}, 45(7-8):455--457, 1927.

\bibitem{einstein1913entwurf}
A~Einstein and M~Grossmann.
\newblock Entwurf einer verallgemeinerten relativit{\"a}tstheorie und einer theorie der gravitation, teubner, leipzig.
\newblock {\em Reprinted in CPAE}, 4, 1913.

\bibitem{einstein1905tragheit}
Albert Einstein.
\newblock Ist die tr{\"a}gheit eines k{\"o}rpers von seinem energieinhalt abh{\"a}ngig?
\newblock {\em Annalen der Physik}, 323(13):639--641, 1905.

\bibitem{einstein1905elektrodynamik}
Albert Einstein.
\newblock Zur elektrodynamik bewegter k{\"o}rper.
\newblock {\em Annalen der physik}, 322(10):891--921, 1905.

\bibitem{einstein1906prinzip}
Albert Einstein.
\newblock Das prinzip von der erhaltung der schwerpunktsbewegung und die tr{\"a}gheit der energie.
\newblock {\em Annalen der Physik}, 325(8):627--633, 1906.

\bibitem{einstein1907relativitatsprinzip}
Albert Einstein.
\newblock {\"U}ber die vom relativit{\"a}tsprinzip geforderte tr{\"a}gheit der energie.
\newblock {\em Annalen der Physik}, 328(7):371--384, 1907.

\bibitem{einstein1911einfluss}
Albert Einstein.
\newblock {\"U}ber den einflu{\ss} der schwerkraft auf die ausbreitung des lichtes.
\newblock {\em Annalen der Physik}, 340(10):898--908, 1911.

\bibitem{einstein1912theorie}
Albert Einstein.
\newblock Zur theorie des statischen gravitationsfeldes.
\newblock {\em Annalen der Physik}, 343(7):443--458, 1912.

\bibitem{einstein1935can}
Albert Einstein, Boris Podolsky, and Nathan Rosen.
\newblock Can quantum-mechanical description of physical reality be considered complete?
\newblock {\em Physical review}, 47(10):777, 1935.

\bibitem{engler2009multiscale}
Adam~J Engler, Patrick~O Humbert, Bernhard Wehrle-Haller, and Valerie~M Weaver.
\newblock Multiscale modeling of form and function.
\newblock {\em Science}, 324(5924):208--212, 2009.

\bibitem{epstein1999agent}
Joshua~M Epstein.
\newblock Agent-based computational models and generative social science.
\newblock {\em Complexity}, 4(5):41--60, 1999.

\bibitem{escande1985stochasticity}
Dominique~F Escande.
\newblock Stochasticity in classical hamiltonian systems: universal aspects.
\newblock {\em Physics Reports}, 121(3-4):165--261, 1985.

\bibitem{bohr1976niels}
Peter Farindon.
\newblock The correspondence principle (1918-1923).
\newblock In J.R. Nielsen and L~Rosenfeld, editors, {\em Niels Bohr Collected Works}, volume~3. North Holland, 1976.

\bibitem{fathi2012social}
Alircza Fathi, Jessica~K Hodgins, and James~M Rehg.
\newblock Social interactions: A first-person perspective.
\newblock In {\em Computer Vision and Pattern Recognition (CVPR), 2012 IEEE Conference on}, pages 1226--1233. IEEE, 2012.

\bibitem{feldman2003reconceptualizing}
Martha~S Feldman and Brian~T Pentland.
\newblock Reconceptualizing organizational routines as a source of flexibility and change.
\newblock {\em Administrative science quarterly}, 48(1):94--118, 2003.

\bibitem{feldman2016beyond}
Martha~S Feldman, Brian~T Pentland, Luciana D'Adderio, and Nathalie Lazaric.
\newblock Beyond routines as things: Introduction to the special issue on routine dynamics, 2016.

\bibitem{fernando1989first}
GW~Fernando, Guo-Xin Qian, M~Weinert, and JW~Davenport.
\newblock First-principles molecular dynamics for metals.
\newblock {\em Physical Review B}, 40(11):7985, 1989.

\bibitem{feynman1965quantum}
RP~Feynman and AR~Hibbs.
\newblock Quantum mechanics and path integration, 1965.

\bibitem{fischer1977hartree}
Charlotte~Froese Fischer.
\newblock Hartree--fock method for atoms. a numerical approach.
\newblock 1977.

\bibitem{fox1988organizational}
Mark~S Fox.
\newblock An organizational view of distributed systems.
\newblock In {\em Readings in Distributed Artificial Intelligence}, pages 140--150. Elsevier, 1988.

\bibitem{frank2003fokker}
TD~Frank, PJ~Beek, and R~Friedrich.
\newblock Fokker-planck perspective on stochastic delay systems: Exact solutions and data analysis of biological systems.
\newblock {\em Physical Review E}, 68(2):021912, 2003.

\bibitem{friedrich1982defense}
Robert~J Friedrich.
\newblock In defense of multiplicative terms in multiple regression equations.
\newblock {\em American Journal of Political Science}, pages 797--833, 1982.

\bibitem{funtowicz1994emergent}
Silvio Funtowicz and Jerome~R Ravetz.
\newblock Emergent complex systems.
\newblock {\em Futures}, 26(6):568--582, 1994.

\bibitem{gasser1993organizations}
Les Gasser, Ingemar Hulthage, Brian Leverich, Jon Lieb, and Ann Majchrzak.
\newblock Organizations as complex, dynamic design problems.
\newblock In {\em Portuguese Conference on Artificial Intelligence}, pages 1--12. Springer, 1993.

\bibitem{gatti2005new}
Domenico~Delli Gatti, Corrado Di~Guilmi, Edoardo Gaffeo, Gianfranco Giulioni, Mauro Gallegati, and Antonio Palestrini.
\newblock A new approach to business fluctuations: heterogeneous interacting agents, scaling laws and financial fragility.
\newblock {\em Journal of Economic behavior \& organization}, 56(4):489--512, 2005.

\bibitem{gell1996information}
Murray Gell-Mann and Seth Lloyd.
\newblock Information measures, effective complexity, and total information.
\newblock {\em Complexity}, 2(1):44--52, 1996.

\bibitem{gershenson2003can}
Carlos Gershenson and Francis Heylighen.
\newblock When can we call a system self-organizing?
\newblock In {\em European Conference on Artificial Life}, pages 606--614. Springer, 2003.

\bibitem{ghabrial2006social}
Amin~S Ghabrial and Mark~A Krasnow.
\newblock Social interactions among epithelial cells during tracheal branching morphogenesis.
\newblock {\em Nature}, 441(7094):746, 2006.

\bibitem{ghenniwa2000interaction}
Hamada Ghenniwa and Mohamed Kamel.
\newblock Interaction devices for coordinating cooperative distributed systems.
\newblock {\em Intelligent Automation \& Soft Computing}, 6(3):173--184, 2000.

\bibitem{gilks1995markov}
Walter~R Gilks, Sylvia Richardson, and David Spiegelhalter.
\newblock {\em Markov chain Monte Carlo in practice}.
\newblock CRC press, 1995.

\bibitem{girard1989towards}
Jean-Yves Girard.
\newblock Towards a geometry of interaction.
\newblock {\em Contemporary Mathematics}, 92(69-108):6, 1989.

\bibitem{glaeser2001measuring}
Edward Glaeser and Jos{\'e} Scheinkman.
\newblock Measuring social interactions.
\newblock {\em Social dynamics}, pages 83--132, 2001.

\bibitem{goldenfeld2018lectures}
Nigel Goldenfeld.
\newblock {\em Lectures on phase transitions and the renormalization group}.
\newblock CRC Press, 2018.

\bibitem{graham1999towards}
John~R Graham and Keith~S Decker.
\newblock Towards a distributed, environment-centered agent framework.
\newblock In {\em International Workshop on Agent Theories, Architectures, and Languages}, pages 290--304. Springer, 1999.

\bibitem{grattan2003joseph}
Ivor Grattan-Guiness and Jerome~R Ravetz.
\newblock Joseph fourier, 1768-1830: A survey of his life and work.
\newblock 2003.

\bibitem{guckenheimer2013nonlinear}
John Guckenheimer and Philip Holmes.
\newblock {\em Nonlinear oscillations, dynamical systems, and bifurcations of vector fields}, volume~42.
\newblock Springer Science \& Business Media, 2013.

\bibitem{gudder2014quantum}
Stanley~P Gudder.
\newblock {\em Quantum probability}.
\newblock Academic Press, 2014.

\bibitem{guimera2005functional}
Roger Guimera and Luis A~Nunes Amaral.
\newblock Functional cartography of complex metabolic networks.
\newblock {\em Nature}, 433(7028):895--900, 2005.

\bibitem{guttal2010social}
Vishwesha Guttal and Iain~D Couzin.
\newblock Social interactions, information use, and the evolution of collective migration.
\newblock {\em Proceedings of the national academy of sciences}, 107(37):16172--16177, 2010.

\bibitem{hammond1994monte}
Brian~L Hammond, William~A Lester, and Peter~James Reynolds.
\newblock {\em Monte Carlo methods in ab initio quantum chemistry}, volume~1.
\newblock World Scientific, 1994.

\bibitem{han1999interacting}
Seung~Kee Han, Tae~Gyu Yim, DE~Postnov, and OV~Sosnovtseva.
\newblock Interacting coherence resonance oscillators.
\newblock {\em Physical Review Letters}, 83(9):1771, 1999.

\bibitem{hassani2009dirac}
Sadri Hassani.
\newblock Dirac delta function.
\newblock In {\em Mathematical methods}, pages 139--170. Springer, 2009.

\bibitem{hesketh2016particle}
Gavin Hesketh.
\newblock {\em The Particle Zoo: The Search for the Fundamental Nature of Reality}.
\newblock Hachette UK, 2016.

\bibitem{hess2002determining}
Berk Hess.
\newblock Determining the shear viscosity of model liquids from molecular dynamics simulations.
\newblock {\em The Journal of chemical physics}, 116(1):209--217, 2002.

\bibitem{hoare1978communicating}
Charles Antony~Richard Hoare.
\newblock Communicating sequential processes.
\newblock {\em Communications of the ACM}, 21(8):666--677, 1978.

\bibitem{hojman1992new}
Sergio~A Hojman.
\newblock A new conservation law constructed without using either lagrangians or hamiltonians.
\newblock {\em Journal of Physics A: Mathematical and General}, 25(7):L291, 1992.

\bibitem{holland1992complex}
John~H Holland.
\newblock Complex adaptive systems.
\newblock {\em Daedalus}, pages 17--30, 1992.

\bibitem{holling2001understanding}
Crawford~S Holling.
\newblock Understanding the complexity of economic, ecological, and social systems.
\newblock {\em Ecosystems}, 4(5):390--405, 2001.

\bibitem{horodecki2009quantum}
Ryszard Horodecki, Pawe{\l} Horodecki, Micha{\l} Horodecki, and Karol Horodecki.
\newblock Quantum entanglement.
\newblock {\em Reviews of modern physics}, 81(2):865, 2009.

\bibitem{howell1992protein}
Nazlin~K Howell.
\newblock Protein-protein interactions.
\newblock In {\em Biochemistry of food proteins}, pages 35--74. Springer, 1992.

\bibitem{huberman2008social}
Bernardo~A Huberman, Daniel~M Romero, and Fang Wu.
\newblock Social networks that matter: Twitter under the microscope.
\newblock {\em arXiv preprint arXiv:0812.1045}, 2008.

\bibitem{hunziker2000quantum}
Walter Hunziker and Israel~Michael Sigal.
\newblock The quantum n-body problem.
\newblock {\em Journal of Mathematical Physics}, 41(6):3448--3510, 2000.

\bibitem{iberall1987physics}
Arthur~S Iberall and Harry Soodak.
\newblock A physics for complex systems.
\newblock In {\em Self-organizing systems}, pages 499--520. Springer, 1987.

\bibitem{ishida1992organization}
Toru Ishida, Les Gasser, and Makoto Yokoo.
\newblock Organization self-design of distributed production systems.
\newblock {\em IEEE Transactions on Knowledge and Data Engineering}, 4(2):123--134, 1992.

\bibitem{itzykson2006quantum}
Claude Itzykson and Jean-Bernard Zuber.
\newblock {\em Quantum field theory}.
\newblock Courier Corporation, 2006.

\bibitem{jammer1989conceptual}
Max Jammer.
\newblock {\em The conceptual development of quantum mechanics}.
\newblock Tomash, 1989.

\bibitem{jurgenson2011evolving}
ED~Jurgenson, P~Navr{\'a}til, and RJ~Furnstahl.
\newblock Evolving nuclear many-body forces with the similarity renormalization group.
\newblock {\em Physical Review C}, 83(3):034301, 2011.

\bibitem{kaiser2014shut}
David Kaiser.
\newblock Shut up and calculate!
\newblock {\em Nature}, 505(7482):153--155, 2014.

\bibitem{katok1997introduction}
Anatole Katok and Boris Hasselblatt.
\newblock {\em Introduction to the modern theory of dynamical systems}, volume~54.
\newblock Cambridge university press, 1997.

\bibitem{kevrekidis2007emergent}
Panayotis~G Kevrekidis, Dimitri~J Frantzeskakis, and Ricardo Carretero-Gonz{\'a}lez.
\newblock {\em Emergent nonlinear phenomena in Bose-Einstein condensates: theory and experiment}, volume~45.
\newblock Springer Science \& Business Media, 2007.

\bibitem{kitano2002computational}
Hiroaki Kitano.
\newblock Computational systems biology.
\newblock {\em Nature}, 420(6912):206, 2002.

\bibitem{ladyman2013complex}
James Ladyman, James Lambert, and Karoline Wiesner.
\newblock What is a complex system?
\newblock {\em European Journal for Philosophy of Science}, 3(1):33--67, 2013.

\bibitem{lakkaraju2008norm}
Kiran Lakkaraju and Les Gasser.
\newblock Norm emergence in complex ambiguous situations.
\newblock In {\em Proceedings of the AAAI workshop on coordination, organizations, institutions and norms AAAI, Chicago}, 2008.

\bibitem{lan2006interplay}
Yueheng Lan and Garegin~A Papoian.
\newblock The interplay between discrete noise and nonlinear chemical kinetics in a signal amplification cascade.
\newblock {\em The Journal of chemical physics}, 125(15):154901, 2006.

\bibitem{lancaster2014quantum}
Tom Lancaster and Stephen~J Blundell.
\newblock {\em Quantum field theory for the gifted amateur}.
\newblock OUP Oxford, 2014.

\bibitem{lee2012bayesian}
Peter~M Lee.
\newblock {\em Bayesian statistics: an introduction}.
\newblock John Wiley \& Sons, 2012.

\bibitem{leibniz1989specimen}
Gottfried~Wilhelm Leibniz.
\newblock Specimen dynamicum.
\newblock In {\em Philosophical Papers and Letters}, pages 435--452. Springer, 1989.

\bibitem{levine2009molecular}
Raphael~D Levine.
\newblock {\em Molecular reaction dynamics}.
\newblock Cambridge University Press, 2009.

\bibitem{lewis1981direct}
Michael Lewis and Candice Feiring.
\newblock Direct and indirect interactions in social relationships.
\newblock {\em Advances in infancy research}, 1981.

\bibitem{li1998influence}
Hong Li and Xianglin Yang.
\newblock The influence of stochastic perturbation on dark soliton-distributed amplification transmission system and its suppression.
\newblock {\em Microwave and Optical Technology Letters}, 17(1):58--62, 1998.

\bibitem{li2001multiscale}
Jinghai Li and Mooson Kwauk.
\newblock Multiscale nature of complex fluid- particle systems.
\newblock {\em Industrial \& Engineering Chemistry Research}, 40(20):4227--4237, 2001.

\bibitem{li2004multi}
Jinghai Li, Jiayuan Zhang, Wei Ge, and Xinhua Liu.
\newblock Multi-scale methodology for complex systems.
\newblock {\em Chemical Engineering Science}, 59(8-9):1687--1700, 2004.

\bibitem{li2012quorum}
Yung-Hua Li and Xiaolin Tian.
\newblock Quorum sensing and bacterial social interactions in biofilms.
\newblock {\em Sensors}, 12(3):2519--2538, 2012.

\bibitem{lichtenberg2013regular}
Allan~J Lichtenberg and Michael~A Lieberman.
\newblock {\em Regular and stochastic motion}, volume~38.
\newblock Springer Science \& Business Media, 2013.

\bibitem{liu2008monte}
Jun~S Liu.
\newblock {\em Monte Carlo strategies in scientific computing}.
\newblock Springer Science \& Business Media, 2008.

\bibitem{lobkovsky2011predictability}
Alexander~E Lobkovsky, Yuri~I Wolf, and Eugene~V Koonin.
\newblock Predictability of evolutionary trajectories in fitness landscapes.
\newblock {\em PLoS computational biology}, 7(12):e1002302, 2011.

\bibitem{loewenstein1981junctional}
WERNER~R Loewenstein.
\newblock Junctional intercellular communication: the cell-to-cell membrane channel.
\newblock {\em Physiological Reviews}, 61(4):829--913, 1981.

\bibitem{luhmann1995social}
Niklas Luhmann.
\newblock {\em Social systems}.
\newblock Stanford University Press, 1995.

\bibitem{luhmann2006system}
Niklas Luhmann.
\newblock System as difference.
\newblock {\em Organization}, 13(1):37--57, 2006.

\bibitem{lutz2002deviation}
Werner~K Lutz, Spyros Vamvakas, Annette Kopp-Schneider, Josef Schlatter, and Helga Stopper.
\newblock Deviation from additivity in mixture toxicity: relevance of nonlinear dose-response relationships and cell line differences in genotoxicity assays with combinations of chemical mutagens and gamma-radiation.
\newblock {\em Environmental health perspectives}, 110(Suppl 6):915, 2002.

\bibitem{macdonald2002incompleteness}
Ranald~R Macdonald.
\newblock The incompleteness of probability models and the resultant implications for theories of statistical inference.
\newblock {\em Understanding Statistics: Statistical Issues in Psychology, Education, and the Social Sciences}, 1(3):167--189, 2002.

\bibitem{mach1907science}
Ernst Mach.
\newblock {\em The science of mechanics: A critical and historical account of its development}.
\newblock Open court publishing Company, 1907.

\bibitem{majda2005information}
Andrew Majda, Rafail~V Abramov, and Marcus~J Grote.
\newblock {\em Information theory and stochastics for multiscale nonlinear systems}, volume~25.
\newblock American Mathematical Soc., 2005.

\bibitem{manski2013identification}
Charles~F Manski.
\newblock Identification of treatment response with social interactions.
\newblock {\em The Econometrics Journal}, 16(1):S1--S23, 2013.

\bibitem{maradudin1963theory}
Alexei~A Maradudin, Elliott~Waters Montroll, George~Herbert Weiss, and IP~Ipatova.
\newblock {\em Theory of lattice dynamics in the harmonic approximation}, volume~3.
\newblock Academic press New York, 1963.

\bibitem{martin2009cell}
Francisco~A Mart{\'i}n, Salvador~C Herrera, and Gin{\'e}s Morata.
\newblock Cell competition, growth and size control in the drosophila wing imaginal disc.
\newblock {\em Development}, 136(22):3747--3756, 2009.

\bibitem{mattia2015supramolecular}
Elio Mattia and Sijbren Otto.
\newblock Supramolecular systems chemistry.
\newblock {\em Nature nanotechnology}, 10(2):111, 2015.

\bibitem{maxwell1904treatise}
James~Clerk Maxwell and Joseph~John Thompson.
\newblock {\em A treatise on electricity and magnetism}, volume~2.
\newblock Clarendon, 1904.

\bibitem{mcclelland1993statistical}
Gary~H McClelland and Charles~M Judd.
\newblock Statistical difficulties of detecting interactions and moderator effects.
\newblock {\em Psychological bulletin}, 114(2):376, 1993.

\bibitem{mehta1963statisticalv}
Madan~Lal Mehta and Freeman~J Dyson.
\newblock Statistical theory of the energy levels of complex systems. v.
\newblock {\em Journal of Mathematical Physics}, 4(5):713--719, 1963.

\bibitem{mendes2002challenge}
Wendy~Berry Mendes, Jim Blascovich, Brian Lickel, and Sarah Hunter.
\newblock Challenge and threat during social interactions with white and black men.
\newblock {\em Personality and Social Psychology Bulletin}, 28(7):939--952, 2002.

\bibitem{miller1988leibniz}
Richard~B Miller.
\newblock Leibniz on the interaction of bodies.
\newblock {\em History of Philosophy Quarterly}, 5(3):245--255, 1988.

\bibitem{milner1993elements}
Robin Milner.
\newblock Elements of interaction: Turing award lecture.
\newblock {\em Communications of the ACM}, 36(1):78--89, 1993.

\bibitem{minker2014foundations}
Jack Minker.
\newblock {\em Foundations of deductive databases and logic programming}.
\newblock Morgan Kaufmann, 2014.

\bibitem{minsky2000law}
Naftaly~H Minsky and Victoria Ungureanu.
\newblock Law-governed interaction: a coordination and control mechanism for heterogeneous distributed systems.
\newblock {\em ACM Transactions on Software Engineering and Methodology (TOSEM)}, 9(3):273--305, 2000.

\bibitem{moffitt2001policy}
Robert~A Moffitt et~al.
\newblock Policy interventions, low-level equilibria, and social interactions.
\newblock {\em Social dynamics}, 4(45-82):6--17, 2001.

\bibitem{moggi1991notions}
Eugenio Moggi.
\newblock Notions of computation and monads.
\newblock {\em Information and computation}, 93(1):55--92, 1991.

\bibitem{nayfeh1995nonlinear}
Ali~H Nayfeh and Char-Ming Chin.
\newblock Nonlinear interactions in a parametrically excited system with widely spaced frequencies.
\newblock {\em Nonlinear Dynamics}, 7(2):195--216, 1995.

\bibitem{nemeth2015sequential}
Christopher Nemeth, Paul Fearnhead, and Lyudmila Mihaylova.
\newblock Sequential monte carlo methods for state and parameter estimation in abruptly changing environments.
\newblock {\em arXiv preprint arXiv:1510.02604}, 2015.

\bibitem{newman2005power}
Mark~EJ Newman.
\newblock Power laws, pareto distributions and zipf's law.
\newblock {\em Contemporary physics}, 46(5):323--351, 2005.

\bibitem{newman2011complex}
Mark~EJ Newman.
\newblock Complex systems: A survey.
\newblock {\em arXiv preprint arXiv:1112.1440}, 2011.

\bibitem{newton1744philosophiae}
Isaac Newton and Edmund Halley.
\newblock {\em Philosophiae naturalis principia mathematica}, volume~62.
\newblock Jussu Societatis Regiae ac typis Josephi Streater, prostant venales apud Sam. Smith, 1744.

\bibitem{nicolis1989physics}
Gregoire Nicolis.
\newblock Physics of far-from-equilibrium systems and self-organisation.
\newblock {\em The new physics}, 11:316--347, 1989.

\bibitem{olivier1989serotonergic}
Berend Olivier, Jan Mos, Jan Van~der Heyden, and Jan Hartog.
\newblock Serotonergic modulation of social interactions in isolated male mice.
\newblock {\em Psychopharmacology}, 97(2):154--156, 1989.

\bibitem{oono1989large}
Yoshitsugu Oono.
\newblock Large deviation and statistical physics.
\newblock {\em Progress of Theoretical Physics Supplement}, 99:165--205, 1989.

\bibitem{oono2012nonlinear}
Yoshitsugu Oono.
\newblock {\em The nonlinear world: conceptual analysis and phenomenology}.
\newblock Springer Science \& Business Media, 2012.

\bibitem{oullier2008social}
Olivier Oullier, Gonzalo~C De~Guzman, Kelly~J Jantzen, Julien Lagarde, and JA~Scott~Kelso.
\newblock Social coordination dynamics: Measuring human bonding.
\newblock {\em Social neuroscience}, 3(2):178--192, 2008.

\bibitem{parnas1985modular}
David~Lorge Parnas, Paul~C Clements, and David~M Weiss.
\newblock The modular structure of complex systems.
\newblock {\em IEEE Transactions on software Engineering}, (3):259--266, 1985.

\bibitem{parr1983density}
Robert~G Parr.
\newblock Density functional theory.
\newblock {\em Annual Review of Physical Chemistry}, 34(1):631--656, 1983.

\bibitem{pattee2012evolving}
Howard~Hunt Pattee.
\newblock Evolving self-reference: matter, symbols, and semantic closure.
\newblock In {\em Laws, language and life}, pages 211--226. Springer, 2012.

\bibitem{penco2006perturbation}
R~Penco and D~Mauro.
\newblock Perturbation theory via feynman diagrams in classical mechanics.
\newblock {\em European journal of physics}, 27(5):1241, 2006.

\bibitem{peskin2018introduction}
Michael~E Peskin.
\newblock {\em An introduction to quantum field theory}.
\newblock CRC Press, 2018.

\bibitem{phillips2005scalable}
James~C Phillips, Rosemary Braun, Wei Wang, James Gumbart, Emad Tajkhorshid, Elizabeth Villa, Christophe Chipot, Robert~D Skeel, Laxmikant Kale, and Klaus Schulten.
\newblock Scalable molecular dynamics with namd.
\newblock {\em Journal of computational chemistry}, 26(16):1781--1802, 2005.

\bibitem{post2009eco}
David~M Post and Eric~P Palkovacs.
\newblock Eco-evolutionary feedbacks in community and ecosystem ecology: interactions between the ecological theatre and the evolutionary play.
\newblock {\em Philosophical Transactions of the Royal Society of London B: Biological Sciences}, 364(1523):1629--1640, 2009.

\bibitem{poston2014catastrophe}
Tim Poston and Ian Stewart.
\newblock {\em Catastrophe theory and its applications}.
\newblock Courier Corporation, 2014.

\bibitem{pryce1998component}
Nat Pryce and Steve Crane.
\newblock Component interaction in distributed systems.
\newblock In {\em Configurable Distributed Systems, 1998. Proceedings. Fourth International Conference on}, pages 71--78. IEEE, 1998.

\bibitem{qu2011multi}
Zhilin Qu, Alan Garfinkel, James~N Weiss, and Melissa Nivala.
\newblock Multi-scale modeling in biology: how to bridge the gaps between scales?
\newblock {\em Progress in biophysics and molecular biology}, 107(1):21--31, 2011.

\bibitem{queller1984kin}
David~C Queller.
\newblock Kin selection and frequency dependence: a game theoretic approach.
\newblock {\em Biological Journal of the Linnean Society}, 23(2-3):133--143, 1984.

\bibitem{raimond2001manipulating}
Jean-Michel Raimond, M~Brune, and Serge Haroche.
\newblock Manipulating quantum entanglement with atoms and photons in a cavity.
\newblock {\em Reviews of Modern Physics}, 73(3):565, 2001.

\bibitem{rajkumar2010cyber}
Ragunathan Rajkumar, Insup Lee, Lui Sha, and John Stankovic.
\newblock Cyber-physical systems: the next computing revolution.
\newblock In {\em Design Automation Conference (DAC), 2010 47th ACM/IEEE}, pages 731--736. IEEE, 2010.

\bibitem{ranganathan2007complexity}
Anand Ranganathan and Roy~H Campbell.
\newblock What is the complexity of a distributed computing system?
\newblock {\em Complexity}, 12(6):37--45, 2007.

\bibitem{rapaport2004art}
Dennis~C Rapaport and Dennis C~Rapaport Rapaport.
\newblock {\em The art of molecular dynamics simulation}.
\newblock Cambridge university press, 2004.

\bibitem{raub1990reputation}
Werner Raub and Jeroen Weesie.
\newblock Reputation and efficiency in social interactions: An example of network effects.
\newblock {\em American Journal of Sociology}, 96(3):626--654, 1990.

\bibitem{raynal1996logical}
Michel Raynal and Mukesh Singhal.
\newblock Logical time: Capturing causality in distributed systems.
\newblock {\em Computer}, 29(2):49--56, 1996.

\bibitem{rieppel2004language}
Olivier Rieppel.
\newblock The language of systematics, and the philosophy of 'total evidence'.
\newblock {\em Systematics and Biodiversity}, 2(1):9--19, 2004.

\bibitem{rybski2009scaling}
Diego Rybski, Sergey~V Buldyrev, Shlomo Havlin, Fredrik Liljeros, and Hern{\'a}n~A Makse.
\newblock Scaling laws of human interaction activity.
\newblock {\em Proceedings of the National Academy of Sciences}, 106(31):12640--12645, 2009.

\bibitem{sadlej1985semi}
Joanna Sadlej.
\newblock {\em Semi-empirical methods of quantum chemistry}.
\newblock Halsted Press, 1985.

\bibitem{sales2007extracting}
Marta Sales-Pardo, Roger Guimera, Andr{\'e}~A Moreira, and Lu{\'i}s A~Nunes Amaral.
\newblock Extracting the hierarchical organization of complex systems.
\newblock {\em Proceedings of the National Academy of Sciences}, 104(39):15224--15229, 2007.

\bibitem{sanchez2005biomimetism}
Cl{\'e}ment Sanchez, Herv{\'e} Arribart, and Marie Madeleine~Giraud Guille.
\newblock Biomimetism and bioinspiration as tools for the design of innovative materials and systems.
\newblock {\em Nature materials}, 4(4):277, 2005.

\bibitem{sawyer2005social}
R~Keith Sawyer.
\newblock {\em Social emergence: Societies as complex systems}.
\newblock Cambridge University Press, 2005.

\bibitem{scheinkman2008social}
Jose~A Scheinkman.
\newblock Social interactions.
\newblock {\em The new palgrave dictionary of economics}, 2, 2008.

\bibitem{schmidt2010renormalization}
Richard Schmidt and Sergej Moroz.
\newblock Renormalization-group study of the four-body problem.
\newblock {\em Physical Review A}, 81(5):052709, 2010.

\bibitem{schrodinger1974life}
Erwin Schr{\"o}dinger.
\newblock {\em What is Life?: The Physical Aspect Og the Living Cell and Mind and Matter}.
\newblock Cambridge University Press, 1974.

\bibitem{scott2003nonlinear}
Alwyn Scott.
\newblock {\em Nonlinear science: emergence and dynamics of coherent structures}.
\newblock Oxford Univ. Press, 2003.

\bibitem{seely2012fractal}
Andrew~JE Seely and Peter Macklem.
\newblock Fractal variability: an emergent property of complex dissipative systems.
\newblock {\em Chaos: An Interdisciplinary Journal of Nonlinear Science}, 22(1):013108, 2012.

\bibitem{seely2014fractal}
Andrew~JE Seely, Kimberley~D Newman, and Christophe~L Herry.
\newblock Fractal structure and entropy production within the central nervous system.
\newblock {\em Entropy}, 16(8):4497--4520, 2014.

\bibitem{segal2003discovering}
Eran Segal, Haidong Wang, and Daphne Koller.
\newblock Discovering molecular pathways from protein interaction and gene expression data.
\newblock {\em Bioinformatics}, 19(suppl\_1):i264--i272, 2003.

\bibitem{shalizi2006methods}
Cosma~Rohilla Shalizi.
\newblock Methods and techniques of complex systems science: An overview.
\newblock In {\em Complex systems science in biomedicine}, pages 33--114. Springer, 2006.

\bibitem{shibata2005noisy}
Tatsuo Shibata and Koichi Fujimoto.
\newblock Noisy signal amplification in ultrasensitive signal transduction.
\newblock {\em Proceedings of the National Academy of Sciences of the United States of America}, 102(2):331--336, 2005.

\bibitem{vsiljak1978decentralized}
DD~{\v{S}}iljak.
\newblock On decentralized control of large scale systems.
\newblock {\em IFAC Proceedings Volumes}, 11(1):1849--1856, 1978.

\bibitem{siljak2011decentralized}
Dragoslav~D Siljak.
\newblock {\em Decentralized control of complex systems}.
\newblock Courier Corporation, 2011.

\bibitem{simon1977organization}
Herbert~A Simon.
\newblock The organization of complex systems.
\newblock In {\em Models of discovery}, pages 245--261. Springer, 1977.

\bibitem{simon1991architecture}
Herbert~A Simon.
\newblock The architecture of complexity.
\newblock In {\em Facets of systems science}, pages 457--476. Springer, 1991.

\bibitem{smolin1995cosmology}
Lee Smolin.
\newblock Cosmology as a problem in critical phenomena.
\newblock In {\em Complex systems and binary networks}, pages 184--223. Springer, 1995.

\bibitem{snijders2002markov}
Tom~AB Snijders.
\newblock Markov chain monte carlo estimation of exponential random graph models.
\newblock {\em Journal of Social Structure}, 3(2):1--40, 2002.

\bibitem{soodak1978homeokinetics}
Harry Soodak and Arthur Iberall.
\newblock Homeokinetics: A physical science for complex systems.
\newblock {\em Science}, 201(4356):579--582, 1978.

\bibitem{sornette2017stock}
Didier Sornette.
\newblock {\em Why stock markets crash: critical events in complex financial systems}.
\newblock Princeton University Press, 2017.

\bibitem{southwood1978substantive}
Kenneth~E Southwood.
\newblock Substantive theory and statistical interaction: Five models.
\newblock {\em American Journal of Sociology}, 83(5):1154--1203, 1978.

\bibitem{stark2000observing}
Jaroslav Stark.
\newblock Observing complexity, seeing simplicity.
\newblock {\em Philosophical Transactions of the Royal Society of London A: Mathematical, Physical and Engineering Sciences}, 358(1765):41--61, 2000.

\bibitem{talcott1997interaction}
Carolyn Talcott.
\newblock Interaction semantics for components of distributed systems.
\newblock In {\em Formal Methods for Open Object-based Distributed Systems}, pages 154--169. Springer, 1997.

\bibitem{thomson2013modern}
Mark Thomson.
\newblock {\em Modern particle physics}.
\newblock Cambridge University Press, 2013.

\bibitem{touchette2009large}
Hugo Touchette.
\newblock The large deviation approach to statistical mechanics.
\newblock {\em Physics Reports}, 478(1-3):1--69, 2009.

\bibitem{turiel2008social}
Elliot Turiel.
\newblock Social decisions, social interactions, and the coordination of diverse judgments.
\newblock {\em Social life and social knowledge: Toward a process account of development}, pages 255--276, 2008.

\bibitem{tzou2007dynamics}
Horn~S Tzou and Lawrence~A Bergman.
\newblock {\em Dynamics and control of distributed systems}.
\newblock Cambridge University Press, 2007.

\bibitem{van1995multiplicative}
Fred~A Van~Eeuwijk.
\newblock Multiplicative interaction in generalized linear models.
\newblock {\em Biometrics}, pages 1017--1032, 1995.

\bibitem{van2008synchronous}
Rob Van~Glabbeek, Ursula Goltz, and Jens-Wolfhard Schicke.
\newblock On synchronous and asynchronous interaction in distributed systems.
\newblock In {\em International Symposium on Mathematical Foundations of Computer Science}, pages 16--35. Springer, 2008.

\bibitem{vespignani2009predicting}
Alessandro Vespignani.
\newblock Predicting the behavior of techno-social systems.
\newblock {\em Science}, 325(5939):425--428, 2009.

\bibitem{vitanyi1997introduction}
Paul~MB Vitanyi and Ming Li.
\newblock {\em An introduction to Kolmogorov complexity and its applications}, volume~34.
\newblock Springer Heidelberg, 1997.

\bibitem{volkov2003oscillatory}
EI~Volkov, E~Ullner, AA~Zaikin, and J~Kurths.
\newblock Oscillatory amplification of stochastic resonance in excitable systems.
\newblock {\em Physical Review E}, 68(2):026214, 2003.

\bibitem{wallace2002constraint}
Mark Wallace.
\newblock Constraint logic programming.
\newblock In {\em Computational logic: Logic programming and beyond}, pages 512--532. Springer, 2002.

\bibitem{walpole2013multiscale}
Joseph Walpole, Jason~A Papin, and Shayn~M Peirce.
\newblock Multiscale computational models of complex biological systems.
\newblock {\em Annual review of biomedical engineering}, 15:137--154, 2013.

\bibitem{wang2011statistical}
Xuefeng Wang, Robert~C Elston, and Xiaofeng Zhu.
\newblock Statistical interaction in human genetics: how should we model it if we are looking for biological interaction?
\newblock {\em Nature Reviews Genetics}, 12(1):74, 2011.

\bibitem{weber2017new}
Edward~P Weber, Denise Lach, and Brent~S Steel.
\newblock {\em New Strategies for Wicked Problems: Science and Solutions in the 21st Century}.
\newblock Oregon State University Press, 2017.

\bibitem{wick1954properties}
Gian-Carlo Wick.
\newblock Properties of bethe-salpeter wave functions.
\newblock {\em Physical Review}, 96(4):1124, 1954.

\bibitem{wieland1997combinatorics}
Thomas Wieland.
\newblock Combinatorics of combinatorial chemistry.
\newblock {\em Journal of Mathematical Chemistry}, 21(2):141--157, 1997.

\bibitem{willetts2001transnational}
Peter Willetts.
\newblock Transnational actors and international organizations in global politics.
\newblock {\em The globalization of world politics}, 2, 2001.

\bibitem{winograd1993categories}
Terry Winograd.
\newblock Categories, disciplines, and social coordination.
\newblock {\em Computer Supported Cooperative Work (CSCW)}, 2(3):191--197, 1993.

\bibitem{woese2002evolution}
Carl~R Woese.
\newblock On the evolution of cells.
\newblock {\em Proceedings of the National Academy of Sciences}, 99(13):8742--8747, 2002.

\bibitem{woolley1977molecular}
RG~Woolley and BT~Sutcliffe.
\newblock Molecular structure and the born—oppenheimer approximation.
\newblock {\em Chemical Physics Letters}, 45(2):393--398, 1977.

\bibitem{wooster2016text}
William~Alfred Wooster.
\newblock {\em A text-book on crystal physics}.
\newblock Cambridge University Press, 2016.

\bibitem{yang2008molecular}
Zhen Yang, Xiaoning Yang, and Zhijun Xu.
\newblock Molecular dynamics simulation of the melting behavior of pt- au nanoparticles with core- shell structure.
\newblock {\em The Journal of Physical Chemistry C}, 112(13):4937--4947, 2008.

\bibitem{zeigler2000theory}
Bernard~P Zeigler, Herbert Praehofer, and Tag~Gon Kim.
\newblock {\em Theory of modeling and simulation: integrating discrete event and continuous complex dynamic systems}.
\newblock Academic press, 2000.

\bibitem{ziliak2008cult}
Steve Ziliak and Deirdre~Nansen McCloskey.
\newblock {\em The cult of statistical significance: How the standard error costs us jobs, justice, and lives}.
\newblock University of Michigan Press, 2008.

\bibitem{zurek1992environment}
Wojciech~H Zurek.
\newblock The environment, decoherence, and the transition from quantum to classical.
\newblock In {\em Quantum Gravity And Cosmology-Proceedings Of The Xxii Gift International Seminar On Theoretical Physics}, page 117. World Scientific, 1992.

\end{thebibliography}

\end{document}